\documentclass[twocolumn]{aastex631}
\usepackage{amsmath}

\def\mbf#1{\mbox{\boldmath ${#1}$}}

\shorttitle{Magnetic bursts and substructures in MRI turbulence}
\shortauthors{Suzuki}
\graphicspath{{./}{figures/}}

\begin{document}

\title{MHD in a cylindrical shearing box II: Intermittent Bursts and Substructures in MRI Turbulence}

\author[0000-0001-9734-9601]{Takeru K. Suzuki}
\affiliation{School of Arts \& Sciences, The University of Tokyo,  
  3-8-1, Komaba, Meguro, Tokyo 153-8902, Japan; stakeru@ea.c.u-tokyo.ac.jp}
\affiliation{Department of Astronomy, The University of Tokyo, 7-3-1,
  Hongo, Bunkyo, Tokyo, 113-0033, Japan}
\affiliation{Komaba Institute for Science, The University of Tokyo, 3-8-1 Komaba, Meguro, Tokyo 153-8902, Japan}




\begin{abstract}
  By performing ideal magnetohydrodynamical (MHD) simulations with weak vertical magnetic fields in unstratified cylindrical shearing boxes with modified boundary treatment, we investigate MHD turbulence excited by magnetorotational instability. 
  The cylindrical simulation exhibits extremely large temporal variation in the magnetic activity compared with the simulation in a normal Cartesian shearing box, although the time-averaged field strengths are comparable in the cylindrical and Cartesian setups. 
  Detailed analysis of the terms describing magnetic-energy evolution with ``triangle diagrams'' surprisingly reveals that in the cylindrical simulation the compression of toroidal magnetic field is unexpectedly as important as the winding due to differential rotation in amplifying magnetic fields and triggering intermittent magnetic bursts, which are not seen in the Cartesian simulation. 
  The importance of the compressible amplification is also true for a cylindrical simulation with tiny curvature; the evolution of magnetic fields in the nearly Cartesian shearing box simulation is fundamentally different from that in the exact Cartesian counterpart.
  The {\it radial gradient of epicyclic frequency}, $\kappa$, which cannot be considered in the normal Cartesian shearing box model, is the cause of this fundamental difference. 
  An additional consequence of the spatial variation of $\kappa$ is continuous and ubiquitous formation of narrow high(low)-density and weak(strong)-field localized structures; 
  seeds of these ring-gap structures are created by the compressible effect and subsequently amplified and maintained under the marginally unstable condition regarding ``viscous-type'' instability.   

\end{abstract}


\section{Introduction} \label{sec:intro}
The local shearing sheet/box model \citep{Goldreich1965,Narayan1987,Hawley1995,Latter2017} is a strong tool to study various physical properties of differentially rotating systems. 
To date, local shearing box simulations have been applied to a variety of astrophysical problems, such as magnetohydrodynamical (MHD hereafter) turbulence \citep[e.g.,][]{Matsumoto1995,Stone1996,Kawazura2022} excited by magnetorotational instability \citep[MRI hereafter; ][]{Velikhov1959,Chandrasekhar1961,Balbus1991}, dynamical and thermal properties of protoplanetary disks \citep[e.g.,][]{Kunz2013,Mori2017,Pucci2021} and accretion disks around compact objects \citep[e.g.][]{Hirose2009,Dempsey2022}, heating accretion disk coronae \citep{Io2014,Bambic2023arxiv}, driving disk winds and outflows \citep{Suzuki2009ApJ,Suzuki2010ApJ,Bai2013,Lesur2013,Fromang2013}, acceleration of non-thermal particles \citep{Hoshino2015,Kimura2016,Bacchini2022}, and amplification of magnetic fields in compact objects \citep{Masada2012ApJ,Guilet2022}.

One of the drawbacks in the normal Cartesian shearing box model is that the positive radial direction is not well defined because the symmetry with respect to the $\pm x$ directions is assumed. As a result, the angular momentum is not defined and the radial accretion flow cannot be properly captured in the Cartesian shearing system. Therefore, mass accretion rate cannot be directly measured in numerical simulations but is inferred from the sum of Maxwell and Reynolds stresses. In addition, the basic physical properties of magnetic-field amplification is considered to be qualitatively different in linear shear flows in Cartesian coordinates and more realistic differentially rotating flows in cylindrical coordinates \citep{Ebrahimi2016MNRAS}. 
Although these problems are not inherent in global simulations \citep[e.g.,][]{Armitage1998,Machida2000,Flock2013,Suzuki2014ApJ,Bethune2017,Takasao2018,Zhu2020,Jacquemin-Ide2021}, 
their numerical cost is generally more expensive. 
Thus, if one can construct a modified model that takes into account global effects in the local approach by breaking the $\pm x$ symmetry, it will be an extremely efficient tool to investigate differentially rotating systems more appropriately. 

To this end, there have been several attempts that try to bridge the local and global concepts over the past few decades \citep[][]{Brandenburg96,Klahr03,Obergaulinger09}.
Along the same lines, \citet[][S19 hereafter]{Suzuki2019PASJ} developed a cylindrical shearing box model by extending the normal Cartesian box model; they applied the conserved quantities of mass, momentum, and magnetic field in cylindrical coordinates to periodic shearing conditions at both radial boundaries.
They demonstrated that quasi-time-steady inward mass accretion is induced by the angular momentum flux that is outwardly transported by MRI-induced turbulence. 

However, S19 performed the only single case with a thin vertical thickness of one scale height. In addition, the treatment for the radial shearing boundaries was not satisfactory; in particular, the difference in the amplitude and frequency of epicyclic perturbations between both ends was causing mismatched fluctuations that travel across the radial boundaries. 
In this paper, we modify the treatment for the radial boundary condition and conduct cylindrical MHD simulations with different ratios between the scale height, $H_0$, and the radial location, $R_0$, in a larger vertical domain.
We examine how the amplification of magnetic fields depends on the curvature effect arising from different aspect ratios, $H_0/R_0$, by comparing the results to those in the exact Cartesian setup.

The structure of the paper is as follows. In Section \ref{sec:model}, we summarize the setup for our simulations in cylindrical shearing boxes. The shearing radial boundary condition is modified from the original method in S19 by separating the shearing variables into the mean and perturbation components, which is described in Appendix \ref{sec:modbd}. Section \ref{sec:res} presents the main results of the simulations, focusing on the transport among the different components of magnetic fields. In Section \ref{sec:dis} we first state the importance of the radial gradient of epicyclic frequency in the time evolution of magnetic fields and  discuss related topics; the detailed formulae for the linear perturbation analyses on viscous-type instability is presented in Appendix \ref{sec:lpa}.
We summarize the paper in Section \ref{sec:sum}.

\section{Cylindrical shearing box simulation}
\label{sec:model}
\subsection{Basic setup}
\label{sec:basics}
We solve ideal MHD equations in cylindrical coordinates, $(R,\phi,z)$, 
\begin{equation}
\frac{d \rho}{d t} + \rho \mbf{\nabla}\cdot \mbf{v} = 0,
\label{eq:mass}
\end{equation}
\begin{equation}
\rho\frac{d\mbf{v}}{dt} = -\mbf{\nabla}\left(p + \frac{B^2}{8\pi}\right) 
+ \left(\frac{\mbf{B}}{4\pi}\cdot \mbf{\nabla}\right)\mbf{B}
-\rho\frac{GM_{\star}}{R^2}\hat{R} 
\label{eq:mom}
\end{equation}
\begin{equation}
\frac{\partial \mbf{B}}{\partial t} = \mbf{\nabla \times (v\times B)}, 
\label{eq:ind}
\end{equation}
and
\begin{equation}
  \mbf{\nabla\cdot B} = 0, 
  \label{eq:divB}
\end{equation}
where $\frac{d}{dt}$ and $\frac{\partial}{\partial t}$ denote Lagrangian and Eulerian time derivatives, respectively; $\rho$, $p$, $\mbf{v}$ and $\mbf{B}$ are density, gas pressure, velocity, and magnetic field, $G$ is the gravitational constant, $M_{\star}$ is the mass of a central star, 
and the ``hat'' stands for a unit vector.

We assume locally isothermal gas with an equation of state,
\begin{equation}
  p=\rho c_{\rm s}^2,
  \label{eq:EoS}
\end{equation}
where $c_{\rm s}$ is isothermal sound speed that depends only on $R$ and does not change with time. 
We employ the following radial dependence of temperature ($\propto c_{\rm s}^2$), 
\begin{equation}
  c_{\rm s}^2 = c_{{\rm s},0}^2\left(\frac{R}{R_0}\right)^{-1},
  \label{eq:Tprfl}
\end{equation}
where the subscript, $0$, indicates that the value is evaluated at the reference radial location, $R=R_0$.

In the numerical simulations, we solve, instead of $\mbf{v}$ in the rest frame, velocity, $\mbf{\delta v} = \mbf{v} - R\Omega_{\rm eq}\hat{\phi}$, or 
\begin{equation}
  (\delta v_R,\delta v_{\phi},\delta v_z) = (v_R,v_{\phi} - R\Omega_{\rm eq},v_z),
  \label{eq:dvphi}
\end{equation}
if expressed in components, evaluated in the frame corotating with the equilibrium rotation frequency, $\Omega_{\rm eq}$.
If we ignore magnetic field and assume $v_R=v_z=0$, the radial component of equation (\ref{eq:mom}) yields
\begin{equation}
  - \frac{v_{\phi,{\rm eq}}^2}{R} = - \frac{GM_{\star}}{R^2} - \frac{1}{\rho}\frac{\partial p}{\partial R}
  \label{eq:rfb}
\end{equation}
under the equilibrium condition. 
Defining $\Omega_{\rm eq}\equiv v_{\phi,{\rm eq}}/R$, from the radial force balance equation (\ref{eq:rfb}) and assuming the radial profile of density, $\rho \propto R^{-\mu}$, we obtain 
\begin{equation}
  \Omega_{\rm eq} = \Omega_{\rm K}\sqrt{1-2f_{\rm K}}, 
  \label{eq:Omgeq}
\end{equation}
where $\Omega_{\rm K} = \sqrt{GM_{\star}/R^3}$ is the Keplerian rotation frequency and
\begin{equation}
  f_{\rm K} = -\frac{1}{\rho}\frac{\partial p}{\partial R}\bigg/2R\Omega_{\rm K}^2 = \frac{(1+\mu)c_{\rm s}^2}{2R^{2} \Omega_{\rm K}^2}
  \label{eq:fK}
\end{equation}
is a sub-Keplerian parameter \citep[][S19]{ada76}. 


Equations (\ref{eq:mass}) -- (\ref{eq:EoS}) are solved with the second-order Godunov + CMoCCT method \citep{Sano99,Evans88,Clarke96}. The azimuthal advection due to $\Omega_{\rm eq}$ (equation \ref{eq:dvphi}) is handled with the ``FARGO (Fast Advection in Rotating Gaseous Objects)''-type algorithm \citep{Masset2000A&AS,Benitez2016ApJS}, which is an improved treatment from S19 to loosen the Courant-Friedrichs-Lewy condition \citep {CFL1928} for the time update.  
The vertical component of the gravity is ignored so that the simulations are performed in vertically unstratified boxes, and thus the periodic boundary condition is applied to the $z$ boundaries as well as the $\phi$ boundaries.

We use the same simulation units as in S19: the physical variables are normalized by the units of $R_0=1$, $\rho_0=1$, and $\Omega_{\rm K,0}=1$; the magnetic field is normalized by $R_0\Omega_{\rm K,0}\sqrt{4\pi\rho_0}$, which cancels the $\sqrt{4\pi}$ factor in the cgs-Gauss units. In the following sections, we conventionally define $2\pi / \Omega_{\rm K,0}$ as ``one rotation time'', which is slightly shorter than the actual time, $2\pi / \Omega_{\rm eq,0}$, it takes for one rotation at $R=R_0$ in the sub-Keplerian condition. 
There are several conserved quantities to verify the numerical treatment in the cylindrical shearing box; see S19 for the details on the other numerical implementation. 

\subsection{Shearing radial boundary condition}
\label{sec:rdshbd}
The radial boundary condition we are adopting is an extension of the shearing periodic condition in a local Cartesian box \citep{Hawley1995}. The basic framework of the shearing radial boundary condition in a local cylindrical box is explained in S19. In short, conserved quantities, instead of primitive variables, are utilized for the boundary treatment by explicitly including the effect of geometrical curvature. 
An improvement from the original prescription in S19 is that we separate the shearing variables, $S$, composed from the conserved quantities, into mean and perturbation components, 
\begin{equation}
  S\mbox{\footnotesize ($R_\pm,\phi,z$)} = \langle S\mbox{\footnotesize ($R_\mp,\phi  \pm\Delta \Omega_{\rm eq}t,z$)}\rangle + \delta S,
  \label{eq:rdbdsum}
\end{equation}
where the subscripts, $-$ and $+$, stand for the inner and outer radial boundaries, $\Delta \Omega_{\rm eq}=\Omega_{\rm eq,-} - \Omega_{\rm eq, +} (>0)$, $\langle \cdots \rangle$ indicates the average over the $\phi - z$ plane, and
\begin{equation}
  \delta S = S - \langle S \rangle
  \label{eq:prtb}
\end{equation}
is the perturbation component (see also Section \ref{sec:ave}).

In the original treatment, $S$ was directly passed to the ghost cell outside the simulation box from the corresponding cells in the simulation domain.
In this prescription, however, as the properties of the disturbances in general differ at the inner and outer boundaries, the mismatched perturbations are exchanged across both boundaries without any correction in an unphysical fashion, which can be most clearly seen in the initial phase of MRI (e.g., Fig. 1 of S19). In order to suppress such contamination across the radial boundaries,  in $\delta S$ we include not only the perturbations from the corresponding sheared cells but also the perturbations on the radially adjacent cell inside the simulation domain.  Additionally, $\delta S$ is rescaled in order that the root-mean-squared (rms) amplitude, $\sqrt{\delta S^2}$, in the ghost cells averaged over the $\phi-z$ surface is matched to the surface averaged $\sqrt{\delta S^2}$ in the adjacent cells inside the simulation domain.
We have free parameters regarding this amplitude matching, which are determined to reproduce steady accretion structure during magnetically inactive periods (see Sections \ref{sec:avsia} and \ref{sec:bdef}).
The detailed numerical implementation for the boundary treatment is described in Appendix \ref{sec:modbd}. 

\subsection{Initial condition}
\label{sec:intcond}
We start the simulations with weak net vertical magnetic field. 
We adopt the same radial dependences of the initial density and vertical field as those employed in S19:
\begin{equation}
  \rho_{\rm init} = \rho_{\rm init,0}\left(\frac{R}{R_0}\right)^{-1}
  \label{eq:rhoinit}
\end{equation}
and
\begin{equation}
  B_{z,{\rm init}} = B_{z,0,{\rm init}}\left(\frac{R}{R_0}\right)^{-1}.
  \label{eq:Bzinit} 
\end{equation}
Equations (\ref{eq:Tprfl}), (\ref{eq:rhoinit}), and (\ref{eq:Bzinit}) guarantee a constant initial plasma $\beta$, and we set
\begin{equation}
  \beta_{z,{\rm init}} = \frac{8\pi\rho_{\rm init} c_{\rm s}^2}{B_{z,{\rm init}}} = 10^4.
  \label{eq:betazinit}
\end{equation}
In order to trigger MRI, we add random velocity perturbations in $v_{R}$ and $v_{\phi}$ to the equilibrium velocity distribution, $(v_R,v_{\phi},v_z) = (0,R\Omega_{\rm eq},0)$.

\subsection{Simulation domain \& resolution}
The simulation domain is a local cylindrical region that covers $(R_- \sim R_+, \phi_- \sim \phi_+, z_- \sim z_+)$.
The azimuthal and vertical spacings, $\Delta \phi$ and $\Delta z$, of each grid cell are constant. The radial grid size, $\Delta R$, is proportional to $R$.

We consider two cases with thin-disk conditions of $c_{\rm s,0} = 0.1R_{0}\Omega_{\rm K,0}$ and $0.01R_{0}\Omega_{\rm K,0}$ at $R=R_0$.
We define the scale height, $H_{0}= c_{\rm s,0}/\Omega_{\rm K,0}$, at  $R=R_0$.
The same box size per $H_0$ is adopted in these two cases, $(L_{R},L_{\phi},L_z) = (R_+-R_-, R_0(\phi_+ - \phi_-), z_+ - z_-) = (4H_0,\frac{5\pi}{3}H_0,4H_0)$, and $(L_R,L_{\phi},L_z)$ is resolved with grid points of $(N_R,N_{\phi},N_z) = (256,320,256)$ (see Table \ref{tab:boxresl} for the summary).
We note that the vertical box size is four times as large as that adopted in S19 to cover the sufficient number of MRI channel-mode wavelengths in the saturated state \citep[see, e.g., ][for discussion on the numerical domain size in Cartesian simulations]{Bodo2008,Johansen2009,Shi2016}. 

\begin{table*}
  \begin{center}
  \begin{tabular}{|c|ccc|ccc|ccc|ccc|}
    \hline
    Model & \multicolumn{9}{c|}{Simulation domain} & \multicolumn{3}{c|}{Resolution} \\
    \hline
    \hline
    Cylindrical box & $R_-$  & $R_+$ & $L_R$ & $\phi_-$ & $\phi_+$ & $L_{\phi}$ & $z_-$ & $z_+$ & $L_z$ & $N_R$ & $N_{\phi}$ & $N_z$ \\
    \hline
    $H_0/R_0 = 0.1$ & $0.82R_0$  & $1.22R_0$ & $4H_0$ & $-\pi/12$ & $\pi/12$ & $(5\pi/3)H_0$ & $-0.2R_0$ & $0.2R_0$ & $4H_0$ & $256$ & $320$ & $256$ \\
    $H_0/R_0 = 0.01$ & $0.980R_0$ & $1.020R_0$ & $4H_0$ & $-\pi/120$ & $\pi/120$ & $(5\pi/3)H_0$ & $-0.02R_0$ & $0.02R_0$ & $4H_0$ & $256$ & $320$ & $256$ \\
    \hline
    \hline
    Cartesian box & $x_-$  & $x_+$ & $L_x$ & $y_-$ & $y_+$ & $L_{y}$ & $z_-$ & $z_+$ & $L_z$ & $N_x$ & $N_y$ & $N_z$ \\
    \hline
    $R_0 \Rightarrow \infty$ & $-2H_0$ & $2H_0$ & $4H_0$ & $-(5\pi/6)H_0$ & $(5\pi/6)H_0$ & $(5\pi/3)H_0$ & $-2H_0$ & $2H_0$ & $4H_0$ & $256$ & $320$ & $256$ \\
    \hline
  \end{tabular}
  \caption{Summary of the simulation domains and resolutions.
  \label{tab:boxresl}}
  \end{center}
\end{table*}

The numerical resolution $=64/H_0$ for the $R$ and $z$ directions is the same as that in S19. It is $\approx 61/H_0$  in the $\phi$ direction, which is slightly higher than $\approx 49/H_0$ in S19. 
For comparison, we also carry out a numerical simulation in a local Cartesian box with the ``same'' box size, $(L_x,L_y,L_z) = (L_R,L_{\phi},L_z)$, per $H_0$ and the same numerical resolution, $(N_x,N_y,N_z) = (N_R,N_{\phi},N_z)$ (Table \ref{tab:boxresl}). 

\subsection{Spatial and temporal average}
\label{sec:ave}
To analyze numerical data, we take various averages of simulated quantities. We express $\langle A \rangle$ as the $\phi$- and $z$- integrated average of a variable, $A$, at $R$: 
\begin{equation}
  \langle A \rangle \equiv
  \frac{\int_{z_{-}}^{z_{+}}\int_{\phi_{-}}^{\phi_{+}} d\phi dz
    A}{\int_{z_{-}}^{z_{+}}\int_{\phi_{-}}^{\phi_{+}} d\phi dz}.
  \label{eq:pzave}
\end{equation}
We define 
\begin{equation}
  [A]_{R_1}^{R_2} = \frac{\int_{R_1}^{R_2} R dR\langle A\rangle}{\int_{R_1}^{R_2} R dR}
  \label{eq:Vave}
\end{equation}
as the volume-integrated average.
For the average over the entire simulation domain, we simply express
\begin{equation}
  [A] \equiv [A]_{R_-}^{R_+}. 
  \label{eq:Vaveall}
\end{equation}
We write
\begin{equation}
  \overline{A} = \frac{\int_{t_1}^{t_2} dt A}{\int_{t_1}^{t_2} dt}
  \label{eq:tave}
\end{equation}
as the average over time between $t_1$ and $t_2$.
In all the simulated cases, the magnetic field grows to the saturated state after $t\gtrsim 20$ rotations. Thus, we take the temporal average from $t_1=25$ rotations to the end of the simulation at $t_2=t_{\rm final} = 300$ rotations.

\subsection{Evolution of magnetic energy}
\label{sec:evB}
\begin{table}
  \begin{center}
    \begin{tabular}{ccc}
      \hline
      & Label & Unabbreviated expression\\
      \hline
      \hline
       & $[\phi\Rightarrow\hspace{-4mm}_{_{\phi}}~\, R]$ & $\left[ \frac{B_R}{4\pi R}\frac{\partial}{\partial \phi}(v_R B_{\phi})\right]/\left[\frac{B_R^2}{8\pi}\right]$\\
      $\frac{\partial\ln\left[{B_R^2}\right]}{\partial t}$ & $[R\Rightarrow\hspace{-4mm}_{_{\phi}}~\, R]$ & $-\left[ \frac{B_R}{4\pi R}\frac{\partial}{\partial \phi}(v_{\phi} B_R)\right]/\left[\frac{B_R^2}{8\pi}\right]$\\
       & $[R\Rightarrow\hspace{-4mm}_{_z}~\, R]$ & $-\left[ \frac{B_R}{4\pi}\frac{\partial}{\partial z}(v_z B_R)\right]/\left[\frac{B_R^2}{8\pi}\right]$\\
       & $[z\Rightarrow\hspace{-4mm}_{_z}~\, R]$ & $\left[ \frac{B_R}{4\pi}\frac{\partial}{\partial z}(v_R B_z)\right]/\left[\frac{B_R^2}{8\pi}\right]$\\
      \hline
       & $[z\Rightarrow\hspace{-4mm}_{_z}~\, \phi]$ & $\left[ \frac{B_\phi}{4\pi}\frac{\partial}{\partial z}(v_\phi B_z)\right]/\left[\frac{B_\phi^2}{8\pi}\right]$\\
      $\frac{\partial \ln \left[ B_{\phi}^2\right]}{\partial t}$ & $[ \phi\Rightarrow\hspace{-4mm}_{_z}~\, \phi]$ & $-\left[ \frac{B_\phi}{4\pi}\frac{\partial}{\partial z}(v_z B_{\phi})\right]/\left[\frac{B_\phi^2}{8\pi}\right]$\\
       &$[\phi\Rightarrow\hspace{-4mm}_{_{_R}}~\, \phi]$ & $-\left[ \frac{B_\phi}{4\pi}\frac{\partial}{\partial R}(v_R B_{\phi})\right]/\left[\frac{B_\phi^2}{8\pi}\right]$\\
       &$[R\Rightarrow\hspace{-4mm}_{_{_R}}~\, \phi]$ & $\left[ \frac{B_\phi}{4\pi}\frac{\partial}{\partial R}(v_\phi B_R)\right]/\left[\frac{B_\phi^2}{8\pi}\right]$\\
      \hline
       &$[R\Rightarrow\hspace{-4mm}_{_{_R}}~\, z]$ & $\left[ \frac{B_z}{4\pi R}\frac{\partial}{\partial R}(Rv_z B_R)\right]/\left[\frac{B_z^2}{8\pi}\right]$\\
      $\frac{\partial \ln \left[B_z^2 \right]}{\partial t}$ & $[z\Rightarrow\hspace{-4mm}_{_{_R}}~\, z]$ & $-\left[ \frac{B_z}{4\pi R}\frac{\partial}{\partial R}(Rv_R B_z)\right]/\left[\frac{B_z^2}{8\pi}\right]$\\
       &$[z\Rightarrow\hspace{-4mm}_{_{\phi}}~\, z]$ & $-\left[\frac{B_z}{4\pi R}\frac{\partial}{\partial \phi}(v_{\phi} B_z)\right]/\left[\frac{B_z^2}{8\pi}\right]$\\
       &$[\phi\Rightarrow\hspace{-4mm}_{_{\phi}}~\, z]$ & $\left[\frac{B_z}{4\pi R}\frac{\partial}{\partial \phi}(v_z B_{\phi})\right]/\left[\frac{B_z^2}{8\pi}\right]$\\
      \hline
    \end{tabular}
  \end{center}
  \caption{Summary of the volume-integrated averages for the evolution of magnetic energy. See equations (\ref{eq:BengR}) -- (\ref{eq:Bengz}) for the detail.  
    \label{tab:dB2}}
\end{table}

Taking the inner product of the induction equation (\ref{eq:ind}) with $\mbf{B}/4\pi$, we have the equation for the evolution of magnetic energy:
\begin{equation}
  \frac{\partial}{\partial t}\left(\frac{\mbf{B}^2}{8\pi}\right) = \frac{\mbf{B}}{4\pi}\mbf{\cdot \nabla \times (v\times B)}.
  \label{eq:Beng}
\end{equation}
For numerical investigation, we rewrite equation (\ref{eq:Beng}) for $B_R^2$, $B_{\phi}^2$, and $B_z^2$ separately in a logarithmic derivative form: 
\begin{align}
  \frac{\partial\ln B_R^2}{\partial t}&= \left(\frac{B_R}{4\pi R}\frac{\partial}{\partial \phi}(v_R B_{\phi} - v_{\phi} B_R) \right. \nonumber \\ &- \left. \frac{B_R}{4\pi}\frac{\partial}{\partial z}(v_z B_R - v_R B_z)\right) \left(\frac{B_R^2}{8\pi}\right)^{-1} \nonumber \\
  &\equiv (\phi\Rightarrow\hspace{-4mm}_{_{\phi}}~\, R) + (R\Rightarrow\hspace{-4mm}_{_{\phi}}~\, R) + (R\Rightarrow\hspace{-4mm}_{_z}~\, R) + (z\Rightarrow\hspace{-4mm}_{_z}~\,  R) \label{eq:BengR}
\end{align}
\begin{align}
  \frac{\partial \ln B_{\phi}^2}{\partial t}&= \left(\frac{B_\phi}{4\pi}\frac{\partial}{\partial z}(v_{\phi} B_z - v_z B_{\phi}) \right. \nonumber \\ &- \left. \frac{B_{\phi}}{4\pi}\frac{\partial}{\partial R}(v_R B_{\phi} - v_{\phi} B_R)\right)\left(\frac{B_{\phi}^2}{8\pi}\right)^{-1} \nonumber \\
  &\equiv ( z\Rightarrow\hspace{-4mm}_{_z}~\, \phi) + ( \phi\Rightarrow\hspace{-4mm}_{_z}~\, \phi) + (\phi\Rightarrow\hspace{-4mm}_{_{_R}}~\, \phi) + (R\Rightarrow\hspace{-4mm}_{_{_{R}}}~\, \phi) \label{eq:Bengp}
\end{align}
\begin{align}
  \frac{\partial \ln B_z^2}{\partial t}&= \left(\frac{B_z}{4\pi R}\frac{\partial}{\partial R}(R(v_z B_R - v_R B_z)) \right. \nonumber \\ &- \left. \frac{B_z}{4\pi R}\frac{\partial}{\partial \phi}(v_{\phi} B_z - v_z B_{\phi}) \right) \left(\frac{B_z^2}{8\pi}\right)^{-1} \nonumber \\
  &\equiv ( R\Rightarrow\hspace{-4mm}_{_{_R}}~\, z) + ( z\Rightarrow\hspace{-4mm}_{_{_R}}~\, z) + ( z\Rightarrow\hspace{-4mm}_{_{\phi}}~\, z) + ( \phi\Rightarrow\hspace{-4mm}_{_{\phi}}~\, z). \label{eq:Bengz}
\end{align}
To examine in detail the evolution of magnetic field, we analyze the $(i\Rightarrow\hspace{-4.mm}_{_k}~\, j)$ terms on the right-hand side of equations (\ref{eq:BengR}) -- (\ref{eq:Bengz}) integrated over the simulation domain (equations \ref{eq:Vave} and \ref{eq:Vaveall}).
We summarize the volume integrated averages, $[i\Rightarrow\hspace{-4.mm}_{_k}~\, j]$, in Table \ref{tab:dB2}.

Similar numerical investigation on the induction equation has been done to understand dynamo-like magnetic evolution \citep{Brandenburg1995ApJ, Davis2010ApJ}.  An essential extension from these works is that we separate the electromotive force, $\mbf{v\times B}$, into the shearing and compressible parts (Section \ref{sec:evolB}). 

\section{Results}
\label{sec:res}
\subsection{Time Evolution}
\label{sec:tevol}

\begin{figure*}
  \includegraphics[width=6.1cm]{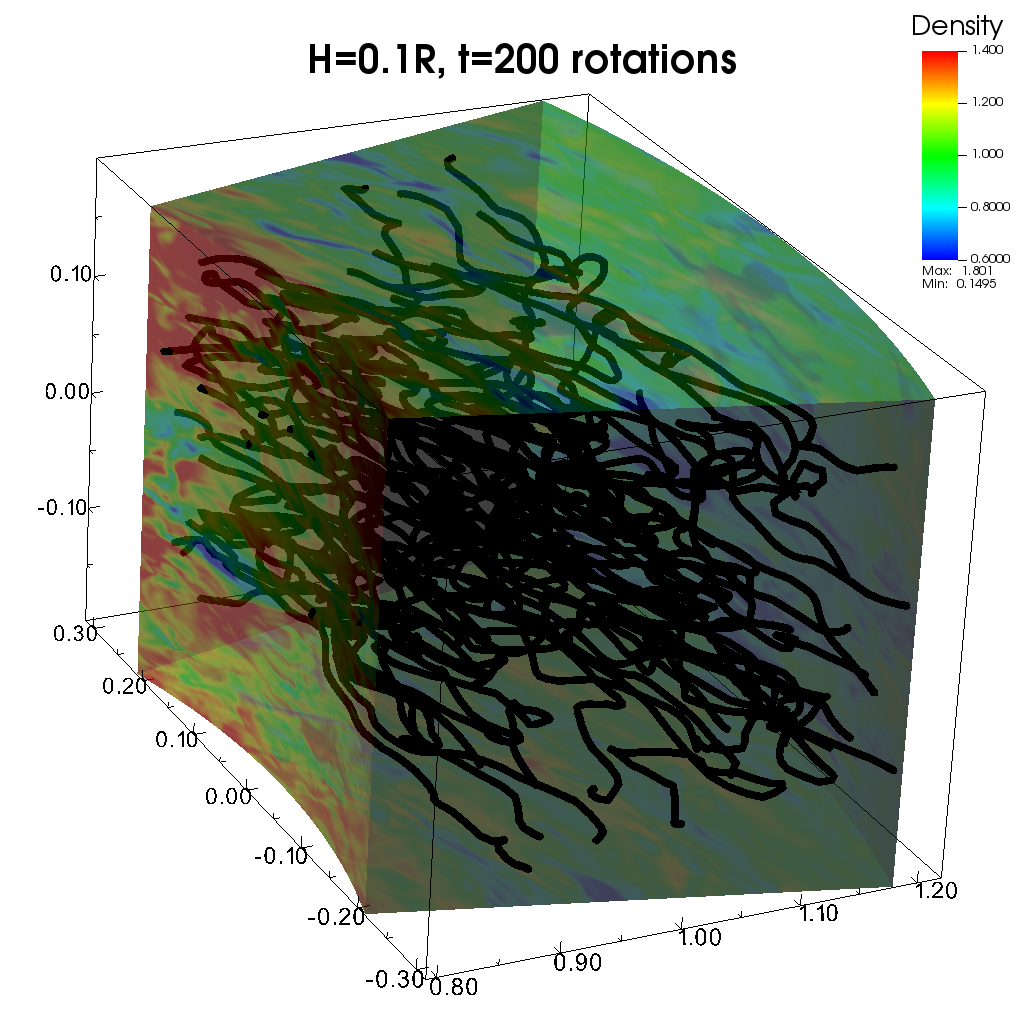}
  \includegraphics[width=5.9cm]{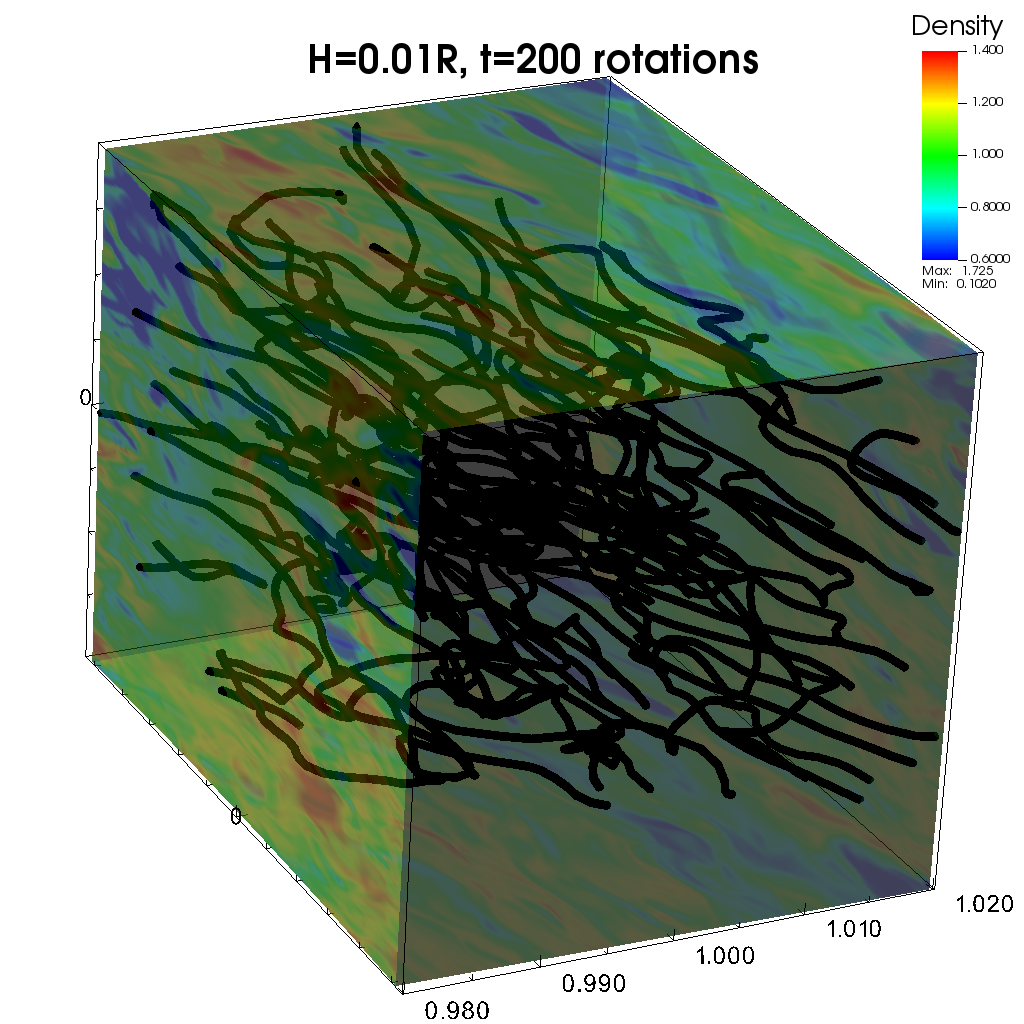}
  \includegraphics[width=6.2cm]{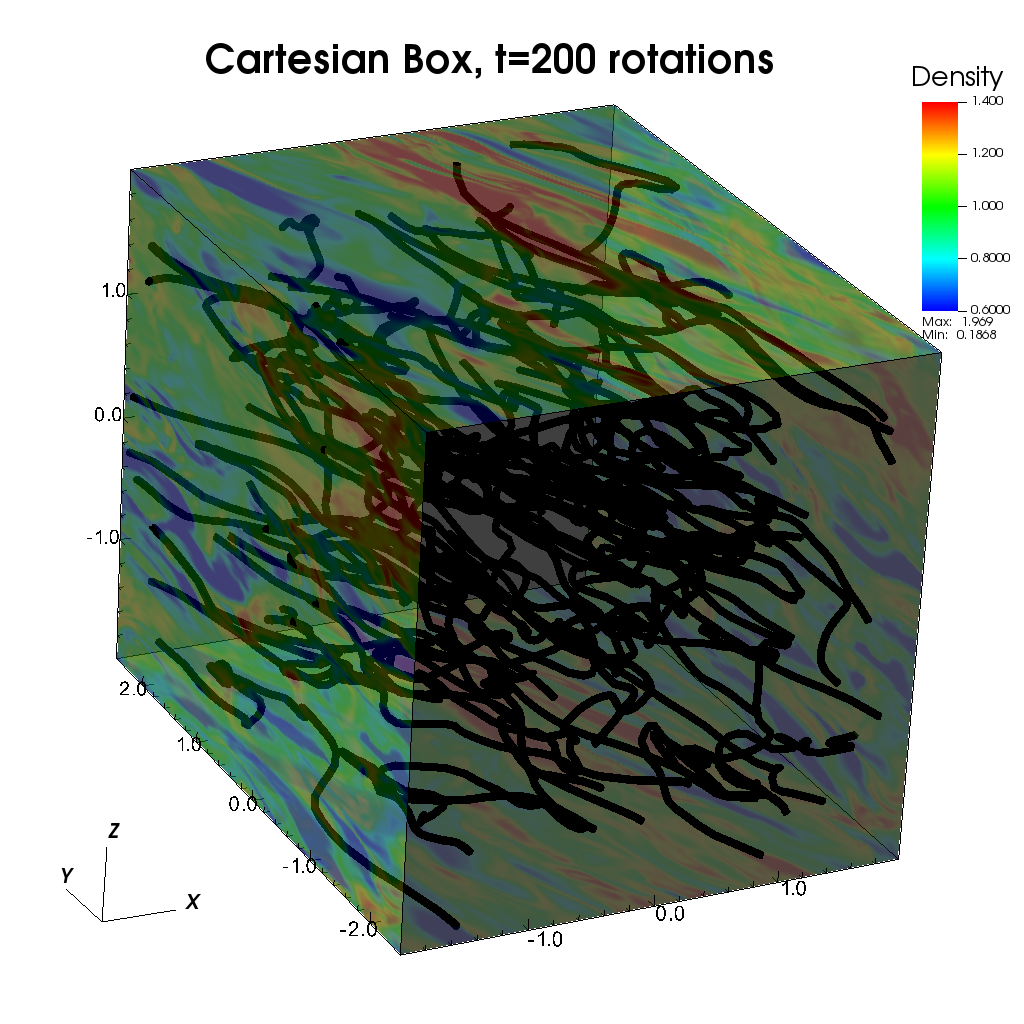}
  \caption{3D views of magnetic field lines (black lines) and density contour (color) of the cylindrical simulations with $H_0/R_0=0.1$ (left) and $0.01$ (middle) and the Cartesian simulation (right) at $t=200$ rotations. Movies are available for the cylindrical case with $H_0/R_0=0.1$ (0-10 rotations and 70-110 rotations) and the Cartesian case (0-30 rotations) at \url{https://ea.c.u-tokyo.ac.jp/astro/Members/stakeru/research/movie/index.html}.
    \label{fig:3Dsnapshots}
  }  
\end{figure*}
We perform numerical simulations of 
the three cases in Table \ref{tab:boxresl} until $t=300$ rotation time.
Figure \ref{fig:3Dsnapshots} presents three-dimensional (3D) snapshots of these cases at $t=200$ rotation time (Movies of the cylindrical case with $H_0/R_0=0.1$ and the Cartesian case are available.).  The three panels illustrate that the magnetic field is turbulent and dominated by the toroidal ($\phi$) component with a moderate level of density fluctuations. 
Table \ref{tab:Bvalues} shows various quantities regarding magneto-turbulence averaged over the time and simulation domain. The 2nd column presents the time- and domain-averaged dimensionless Maxwell stress,
\begin{equation}
  \overline{[\alpha_{\rm M}]} = -\frac{\overline{[B_R B_{\phi}]}}{4\pi \overline{[p]}},  
\end{equation}
and the 3rd --7th columns represent different components of the magnetic energies and their ratios.
While the three cases give similar 
magnetic properties, by a close look at the table, one can recognize that the two cylindrical cases give slightly larger ratios of the poloidal ($R$ and $z$) to toroidal magnetic fields (6th and 7th columns).

The 8th column of Table \ref{tab:Bvalues} shows the time- and volume-averaged Reynolds stress: 
  \begin{equation}
    \overline{[\alpha_{\rm R}]} = \frac{\overline{[\rho v_R \delta v_{\phi}]}}{4\pi \overline{[p]}}.
    \label{eq:Reynolds}
  \end{equation}  
  We find that the cylindrical cases yield much lower $[\alpha_{\rm R}]$ than the Cartesial case by nearly an order of magnitutde. We infer that this large difference between the cylindrical and Cartesian cases is due to the substantially different channels for the magnetic field amplification as will be discussed in Section \ref{sec:evolB}. However, we do not suppose that the quantitative values of $[\alpha_{\rm R}]$ of the cylindrical cases are reliable. 
  This is because it is not clear whether $\delta v_{\phi}$ defined as the deviation from the sub-Keprerian equilibrium rotation, $R \Omega_{\rm eq}$ (equation \ref{eq:dvphi}), is physically reasonable or not to estimate the Reynolds stress when the background rotation profile is deviated from $\Omega_{\rm eq}$ \citep[][see also Section \ref{sec:avsia} for further discussion]{Hawley2001}. For example, if we calculate $[\alpha_{\rm R}]$ by defining $\delta v_{\phi}$ 
  as the deviation from the Keplerian rotation instead of $R \Omega_{\rm eq}$,  then the Reynolds stresses obtained from both cylindrical cases are larger by $\sim 10^{-3}$ than those in Table \ref{tab:Bvalues}. 

\begin{table*}
  \begin{center}
    \begin{tabular}{cccccccc}
      \hline
      Model & $\overline{[\alpha_{\rm M}]}$  & $\overline{[B_R^2]} / 8\pi\overline{[ p ]}$ & $\overline{[B_{\phi}^2]} / 8\pi\overline{[ p ]}$ & $\overline{[B_z^2]} / 8\pi\overline{[ p ]}$ & $\overline{[B_R^2]} / \overline{[B_{\phi}^2]}$ & $\overline{[B_z^2]} / \overline{[B_{\phi}^2]}$ & $\overline{[\alpha_{\rm R}]}$ \\
      \hline
      \hline
      $H_0/R_0 = 0.1$ & $6.22\times 10^{-2}$ & $2.78 \times 10^{-2}$ & $1.11\times 10^{-1}$ & $1.25\times 10^{-2}$ & $0.250$ & $0.112$ & $3.36\times 10^{-3}$ \\
      $H_0/R_0 = 0.01$ & $5.60\times 10^{-2}$ & $2.51\times 10^{-2}$ & $1.00\times 10^{-1}$ & $1.14\times 10^{-2}$ & $0.250$ & $0.113$ & $2.10\times 10^{-3}$\\
      Cartesian Box & $7.16\times 10^{-2} $ & $2.64\times 10^{-2}$ & $1.29\times 10^{-1}$ & $1.32\times 10^{-2}$ & $0.204$ & $0.102$ & $1.56\times 10^{-2}$\\
      \hline
    \end{tabular}
    \caption{Various dimensionless magnetic quantities averaged over $t=25-300$ rotations and the entire simulation domain. The 2nd column shows Maxwell stress; the 3rd, 4th, and 5th columns present the radial, azimuthal and vertical components of magnetic energy normalized by the gas pressure. The 6th and 7th columns show the ratio of the poloidal components of magnetic energy to the azimuthal component of magnetic energy. The 8th column presents Reynolds stress.
    \label{tab:Bvalues}
    }
  \end{center}
\end{table*}

\begin{figure}
  \begin{center}
    \hspace{-1cm}\includegraphics[height=6.6cm]{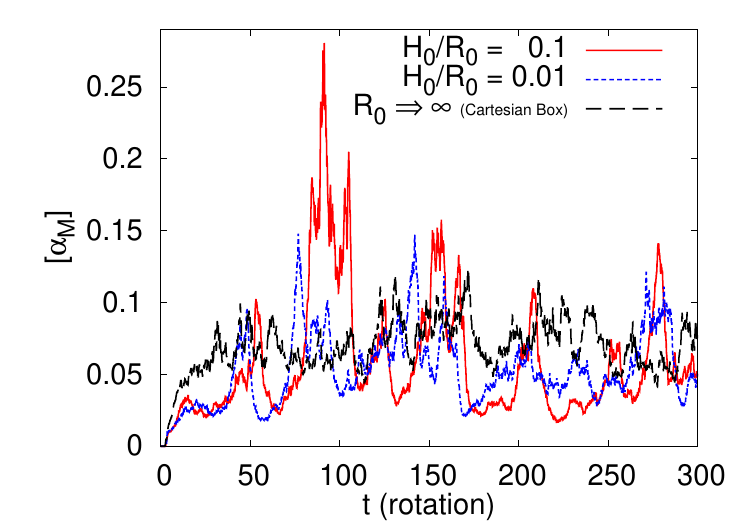}
  \end{center}
  \caption{Comparison of the time evolutions of $[\alpha_{\rm M}]$ averaged over the simulation box for the cylindrical cases with $H_0/R_0=0.1$ (red solid) and $0.01$ (blue dotted) and the Cartesian case (black dashed). 
  \label{fig:alphaM_tevol}}
\end{figure}

Figure \ref{fig:alphaM_tevol} presents the time evolution of the domain-averaged dimensionless Maxwell stress. 
The cylindrical case with $H_0/R_0=0.1$ (red solid) exhibits very large temporal variability; $[\alpha_{\rm M}]$ reaches nearly $0.3$ in the active period between $80 \lesssim t \lesssim 105$ rotations, while $[\alpha_{\rm M}] < 0.05$ is considerably small in the inactive phases.
In contrast, the Cartesian case (black dashed) gives a more time-steady evolution with $0.05\lesssim [\alpha_{\rm M}]\lesssim 0.1$ for most of the simulation time. 
The case with $H_0/R_0=0.01$ (blue dotted) shows intermediate behavior between these two cases.
In spite of the different evolutionary properties, the time-averaged $\overline{[\alpha_{\rm M}]}$'s of these three cases are similar (Table \ref{tab:Bvalues}) because larger $[\alpha_{\rm M}]$ in the active phases and smaller $[\alpha_{\rm M}]$ in the inactive phases are obtained in higher-variability cases.

\subsection{Evolution of magnetic field}
\label{sec:evolB}
\begin{figure*}
   \begin{center}
     \includegraphics[width=9cm]{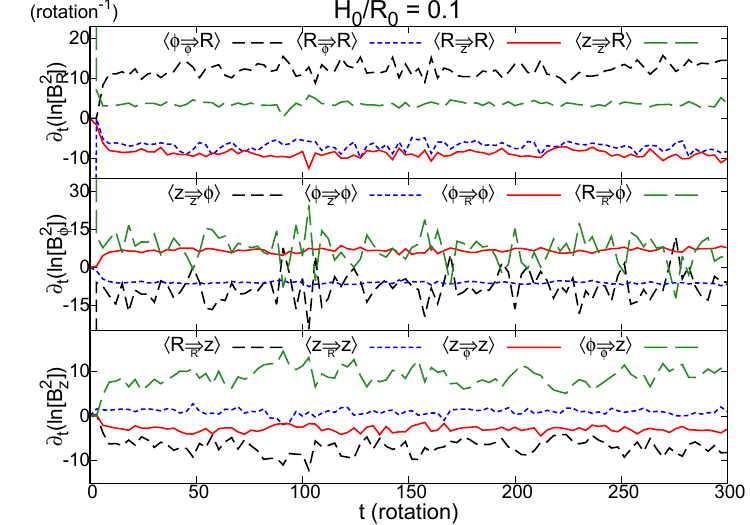}
     \hspace{-0.5cm}
     \includegraphics[width=9cm]{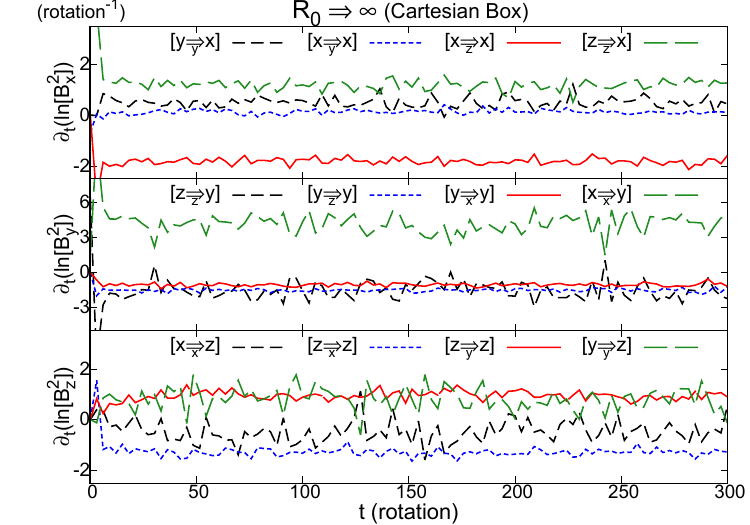}
   \end{center}
   \caption{Comparison of the volume-integrated averages of the variation rates in magnetic energy (equations \ref{eq:BengR} -- \ref{eq:Bengz} and Table \ref{tab:dB2}) of the cylindrical case with $H_0/R_0 = 0.1$ (left) and the Cartesian case (right). What are shown from top to bottom are $\frac{\partial \ln [B_R]^2}{\partial t}$ (or $\frac{\partial \ln [B_x]^2}{\partial t}$),  $\frac{\partial \ln [B_{\phi}]^2}{\partial t}$ (or $\frac{\partial \ln [B_y]^2}{\partial t}$), and $\frac{\partial \ln [B_z]^2}{\partial t}$ in units of (rotation time)$^{-1}$. Smoothing with spline interpolation is applied to the plotted lines to display the long-time evolution. By this treatment, the original numerical data are averaged over $\approx 3$ rotation times.
     \label{fig:dB2tot}
   }
\end{figure*}

\begin{figure*}
  \begin{center}
    \includegraphics[height=7cm]{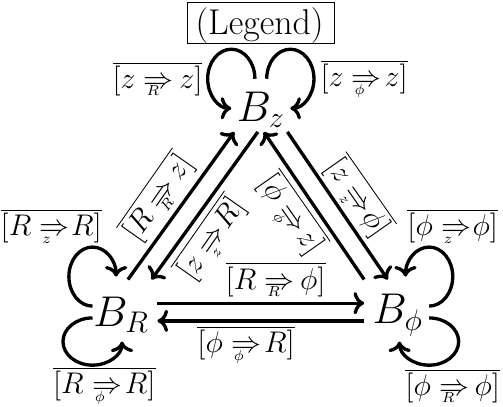}
    \includegraphics[height=7cm]{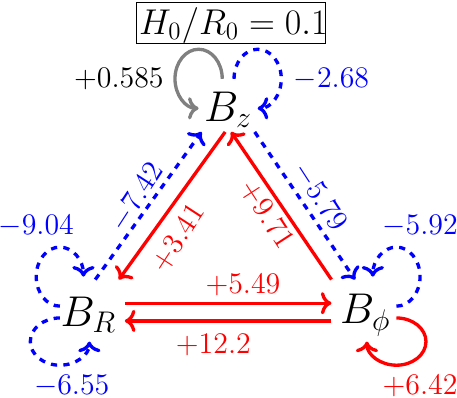}\\
    \vspace{0.6cm}
    \includegraphics[height=7cm]{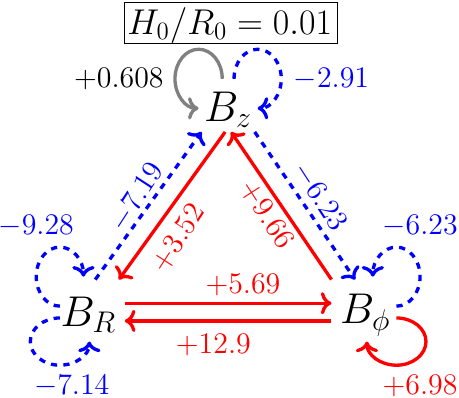}
    \includegraphics[height=7cm]{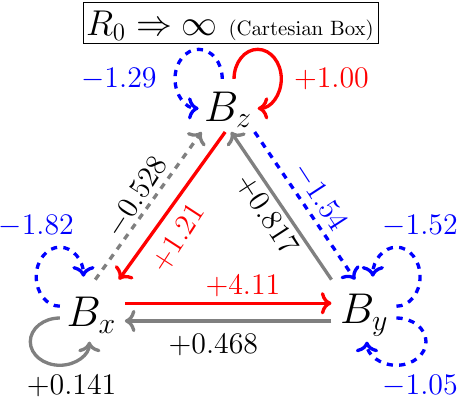}\\
  \end{center}
  \caption{Relation among the three components of magnetic field.
    The top-left triangle illustrates the physical quantities describing each arrow.
    The top-right, bottom-left, and bottom-right triangles show the results of the cylindrical cases with $H_0/R_0=0.1$ and $0.01$, and the Cartesian case, respectively.
    The numerical values indicate the rate of the change of magnetic energy, $\frac{\partial \ln \overline{[B_i^2]}}{\partial t}$, per rotation time averaged over 25 -- 300 rotations (See Table \ref{tab:dB2}).
    The positive (negative) rates with $\ge 1$($\le -1$) are shown by red solid (blue dashed) arrows.
    We note that the sum of the values on the arrows entering each $B_i$ is $=0$ when the time-steady condition is achieved. 
    \label{fig:Triangle01}
    }
\end{figure*}

\begin{figure}
  \begin{center}
    \includegraphics[width=5.4cm]{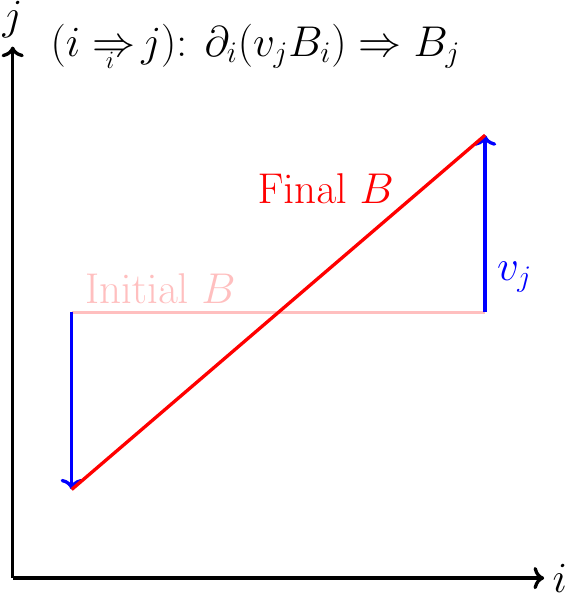}
    \hspace{1cm}
    \includegraphics[width=5.4cm]{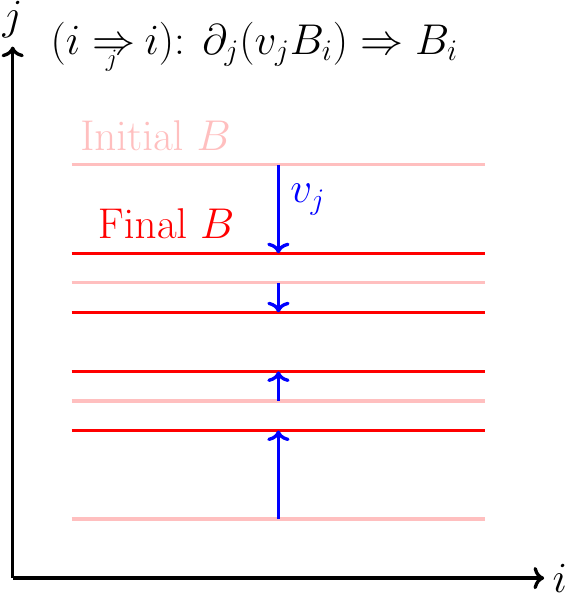}
  \end{center}
  \caption{{\it Left}: The shearing term, $[i\Rightarrow\hspace{-4.mm}_{_i}~\, j]$, stands for the generation (or decay) of $B_j$ from $B_i$ by the $i$ derivative of $v_j$. {\it Right}: The compressible term, $[i\Rightarrow\hspace{-4.mm}_{_j}~\, i]$, represents the convergent amplification (or divergent disperse) of $B_i$ by the $j$ derivative of $v_j$.
  \label{fig:schpic}}
\end{figure}

We have shown that the cylindrical simulations exhibit considerably different temporal magnetic activity from the Cartesian simulation.
In order to understand the difference, we examine the volume averaged variation rates of the magnetic energy (equations \ref{eq:BengR} -- \ref{eq:Bengz} and Table \ref{tab:dB2} in Section \ref{sec:evB}) of the cylindrical case with $H_0/R_0=0.1$ and the Cartesian case in Figure \ref{fig:dB2tot}.
Although some terms, e.g., $[z\Rightarrow\hspace{-4.2mm}_{_z}~\, \phi]$ and $[R\Rightarrow\hspace{-4.2mm}_{_{_R}}~\, \phi]$ of the cylindrical case and $[x\Rightarrow\hspace{-4.2mm}_{_x}~\, z]$ and $[y\Rightarrow\hspace{-4.2mm}_{_y}~\, z]$ in the Cartesian case, exhibit relatively large time variability, most of the terms show rather time-steady behavior;  at least the sign of each term is almost unchanged during the simulation time.

We illustrate all these $[i\Rightarrow\hspace{-4.mm}_{_k}~\, j]$ terms averaged between $t=25$ and $300$ rotations (equation (\ref{eq:tave})) in ``triangle diagrams'' (Figure \ref{fig:Triangle01}), where the result of the cylindrical case with $H_0/R_0=0.01$ is also presented here. We note that, 
when the steady-state condition is satisfied, the sum of numerical values from all four arrows entering each $B_i$ is $0$. Figure \ref{fig:Triangle01} indicates that, while this is approximately satisfied in these three cases, the sums yield tiny positive values. For example, the cylindrical case with $H_0/R_0=0.1$ gives $\partial_t \ln [B_R^2]=0.028$, $\partial_t \ln [B_{\phi}^2]=0.21$, and $\partial_t \ln [B_z^2]=0.20$ (rotation$^{-1}$). These values are much larger than those expected from the small increase in $B_i^2$ during the time averaged period between $t=25$ and 300 rotations (Figure \ref{fig:alphaM_tevol}). Therefore, the net increase in $B_i^2$ by the positive $\partial_t \ln [B_i^2] $ is considered to be compensated by the numerical dissipation of the magnetic fields in a sub-grid scale; in other words, this analysis can quantify the numerical dissipation of magnetic fields in ideal MHD simulations (see also Section \ref{sec:substructure} for discussion on the magnetic diffusion in the ideal MHD condition). 

The variation rates of the two cylindrical cases are similar to each other, but they are very different from those in the Cartesian case (Figure \ref{fig:Triangle01}). This indicates that the physical properties of the magnetic evolution in the local Cartesian box is fundamentally different even from those in the nearly Cartesian box ($H_0/R_0=0.01$). We speculate that this is because the Cartesian shearing box approximation cannot consider the radial variation of epicyclic frequency (see Section \ref{sec:rdepkapp}).
Additionally, the magnitudes of the variation rates are systematically larger in the cylindrical simulations. This is expected to cause the larger temporal variability observed in the cylindrical simulations (Figure \ref{fig:alphaM_tevol}). 

\begin{figure}
  \begin{center}
    \includegraphics[width=6cm]{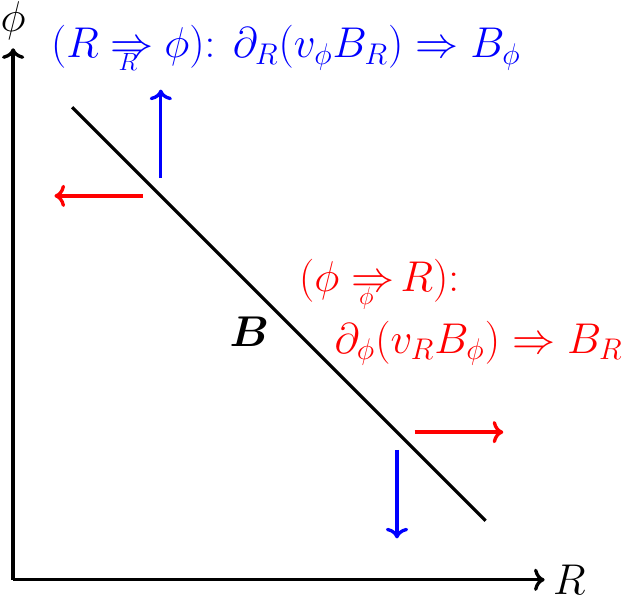}
  \end{center}
  \caption{An example for the amplification of a magnetic field by shearing terms. The magnetic field is distributed with $-B_R B_{\phi}  >0$, which is the usual situation in the system with inner-fast differential rotation.  $B_{\phi}$ increases by radial differential rotation of $v_{\phi}$, which is represented by $(R\Rightarrow\hspace{-4mm}_{_{_R}}~\, \phi)$ (blue). $B_R$ increases by the inward (outward) motion of the inner (outer) region as a result of the outward transport of angular momentum, which is represented by $(\phi\Rightarrow\hspace{-4mm}_{_{\phi}}~\, R)$ (red). 
  \label{fig:sh+AM}}
\end{figure}

We examine each $\overline{[i\Rightarrow\hspace{-4.mm}_{_k}~\, j]}$ term in more detail, particularly focusing on the similarities and differences between the cylindrical and Cartesian simulations.
Let us begin with the $\overline{[i\Rightarrow\hspace{-4.mm}_{_i}~\, j]}$ terms located on the sides of the triangle, which stand for the transfer of magnetic fields between different components of $\mbf{B}$; the physical meaning is the growth or decay of magnetic fields by shearing motion (left panel of Figure \ref{fig:schpic}). 
Although the numerical value of each shearing term is considerably different for the cylindrical and Cartesian simulations, the signs are the same. Thus, the physical properties on the shearing amplification (and attenuation for the negative values) of the magnetic fields are similar at least in a qualitative sense.
Positive $\overline{[z\Rightarrow\hspace{-4mm}_{_z}~\, R]}$ and $\overline{[R\Rightarrow\hspace{-4mm}_{_{_R}}~\, \phi]}$ reflect the standard pathway triggered by MRI \citep[e.g.,][]{Brandenburg1995ApJ,Balbus1998};  $B_R$ is generated from $B_z$ by the MRI, and eventually, $B_{\phi}$ is amplified from the fluctuating $B_R$ by the winding due to the radial differential rotation (blue arrows in Figure \ref{fig:sh+AM}).

One can find that $\overline{[\phi\Rightarrow\hspace{-4mm}_{_{\phi}}~\, R]}$ is also positive.  
This is due to the transfer of angular momentum. The magnetic fields are distributed preferentially along the $-r\phi$ direction by radial differential rotation as shown in Figure \ref{fig:3Dsnapshots}. In stronger-field regions, the angular momentum is more effectively transported outwardly along the field lines. This causes the inner (outer) fluid parcel to move inward (outward), whereas the volume averaged bulk flow keeps the steady accretion structure (see Section \ref{sec:avsia}). Therefore, $B_R$ is generated from $B_{\phi}$
(red arrows in Figure \ref{fig:sh+AM}), namely positive $\overline{[\phi\Rightarrow\hspace{-4mm}_{_{\phi}}~\, R]}$ is obtained. 
Moreover, $\overline{[\phi\Rightarrow\hspace{-4mm}_{_{\phi}}~\, R]}$ ($=+12.2$ and $+12.9$)  in the cylindrical simulations are much larger than $\overline{[y\Rightarrow\hspace{-4mm}_{_y}~\, x]}(=+0.468)$ in the Cartesian simulation. The difference is probably because the angular momentum cannot be defined in the Cartesian setup owing to the assumed symmetry with respect to the $\pm x$ directions (See Section \ref{sec:intro}). 
The leakage from the dominant component of $B_{\phi}$ ($B_y$) to $B_z$ also gives positive $\overline{ [\phi\Rightarrow\hspace{-4mm}_{_{\phi}}~\, z]}$ ($\overline{ [y\Rightarrow\hspace{-4mm}_{_y}~\, z]}$) with the similar tendency concerning the difference between the cylindrical and Cartesian cases. 

Next, we inspect the compressible terms, $\overline{[i\Rightarrow\hspace{-4.mm}_{_j}~\, i]}$, which correspond to the round arrows around each $B_i$ in Figure \ref{fig:Triangle01}. The physical meaning is the amplification (attenuation) of magnetic field by convergent (divergent) flows (right panel of Figure \ref{fig:schpic}). One can see that the signs of some $\overline{[i\Rightarrow\hspace{-4.mm}_{_j}~\, i]}$ terms are different in the cylindrical and Cartesian simulations.
Most remarkable difference is found in the compressible amplification of $B_{\phi}$ along $R$; $\overline{[\phi\Rightarrow\hspace{-4mm}_{_{_R}}~\, \phi]}$ in the cylindrical cases is quite large ($=+6.42$ for $H_0/R_0=0.1$ and $+6.98$ for $H_0/R_0=0.01$), which is in contrast to negative $\overline{[y\Rightarrow\hspace{-4mm}_{_x}~\, y]}(=-1.05)$ in the Cartesian case. Surprisingly, the amplification rate by $\overline{[\phi\Rightarrow\hspace{-4mm}_{_{_R}}~\, \phi]}$ is slightly larger than that of $\overline{[R\Rightarrow\hspace{-4mm}_{_{_R}}~\, \phi]}$, the winding term by radial differential rotation. In other words, the contribution from the compressible flows dominates that from the shear flows in the amplification of $B_{\phi}$ in the realistic cylindrical simulations, which is fundamentally different from the result obtained from the Cartesian simulation. We again infer that this is because of the radial variation of epicyclic frequency, which cannot be considered in the Cartesian setup.
The relative importance of the compression against the shear is a possible and probably plausible reason for the small Reynolds stress in the cylindrical simulations (Section \ref{sec:tevol}). From geometrical considerations, the Reynolds stress $\propto v_R \delta v_{\phi}$ is expected to be generated by shearing motions. If the compressible effect dominates, the sheared stress will be perturbed or interrupted, which will reduce the Reynolds stress.

$\overline{ [R\Rightarrow\hspace{-4mm}_{_{\phi}}~\, R]}$ is also different between the cylindrical and Cartesian simulations. Although $\overline{ [x\Rightarrow\hspace{-4mm}_{_{y}}~\, x]}$ is nearly 0 in the Cartesian simulation, $\overline{ [R\Rightarrow\hspace{-4mm}_{_{\phi}}~\, R]}$ takes a relatively large negative value in the cylindrical simulations, which partially cancels positive $\overline{ [\phi\Rightarrow\hspace{-4mm}_{_{\phi}}~\, R]}$ due to the outward transport of angular momentum, described above (Figure \ref{fig:sh+AM}).

The signs of the compressible terms regarding $B_z$ are all opposite between the cylindrical and Cartesian simulations, whereas the variation rates themselves are not so large.
The radial gradient of epicyclic frequency is a key for the positive $\overline{[z\Rightarrow\hspace{-4mm}_{_{_R}}~\, z]}$ in the cylindrical simulations, which will be discussed in Section \ref{sec:rdepkapp}.

\subsection{Onset of enhanced magnetic activity}
\label{sec:onset}

\begin{figure*}
   \begin{center}
     \includegraphics[width=11cm]{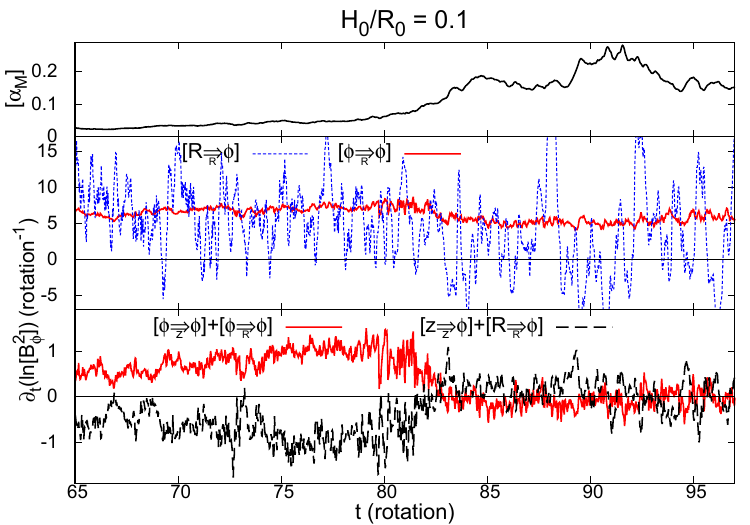}
   \end{center}
   \caption{Time evolution of Maxwell stress (top) and variation rates of $[B_{\phi}^2]$ (middle and bottom) averaged over 
     the domain in the rising and saturated phases of the most active period in the cylindrical simulation with $H_0/R_0 = 0.1$. The middle panel compares the radial compression (red solid) and shear (blue dotted) terms. The bottom panel presents the total compression (red solid) and shear (black dashed) terms. 
     \label{fig:dBp2_B2}
   }
\end{figure*}

We investigate in detail the mechanism of the large temporal magnetic variability (Figure \ref{fig:alphaM_tevol}) observed in the cylindrical case with 
$H_0/R_0=0.1$ by utilizing $\overline{[i\Rightarrow\hspace{-4.mm}_{_k}~\, j]}$ terms.
In this subsection, we scrutinize the rising phase of magnetic activity 
Figure \ref{fig:dBp2_B2} shows the time evolution of 
$[\alpha_{\rm M}]$ (top panel) and variation rates of the most dominate component of the toroidal field, 
$\partial \ln ([B_{\phi}^2])/\partial t$, (middle and bottom panels) during $t=65 - 97$ rotation time, which covers the rising and saturated phases of the largest peak in $[\alpha_{\rm M}]$ (Figure \ref{fig:alphaM_tevol}).

The middle panel of Figure \ref{fig:dBp2_B2} compares the variation rates of $B_{\phi}^2$ due to the radial compression, 
$[\phi\Rightarrow\hspace{-4mm}_{_{_R}}~\, \phi]$ (red solid) and the winding by radial shear, 
$[R\Rightarrow\hspace{-4.5mm}_{_{_R}}~\, \phi]$ (blue dotted).
While the radial shear term greatly fluctuates from negative to positive values with time, the compressible term takes positive values in a more stable manner. In particular, the compressible amplification plays an indispensable role in the growth of $B_{\phi}^2$ in the rising phase of the magnetic activity when the radial shear term stays at a relatively low level. 

The bottom panel of Figure \ref{fig:dBp2_B2} shows the sum of the compressible terms, 
$\left\{[\phi\Rightarrow\hspace{-4mm}_{_z}~\, \phi] + [\phi\Rightarrow\hspace{-4mm}_{_{_R}}~\, \phi]\right\}$ (red solid) and the shearing terms, 
$\left\{[z\Rightarrow\hspace{-4mm}_{_z}~\, \phi] + [R\Rightarrow\hspace{-4.5mm}_{_{_R}}~\, \phi]\right\}$, (black dashed). 
As shown in Figures \ref{fig:dB2tot} \& \ref{fig:Triangle01}, the vertical compression term, $[\phi\Rightarrow\hspace{-4mm}_{_z}~\, \phi]$, is negative, and thus, the variation rate by the total compression is reduced from that by the only radial term (red line in the middle panel). However, the effect of the total compression (red line in the bottom panel) still keeps positive in the initial rising phase of $t\lesssim 83$ rotation time; the compressible amplification plays an essential role in triggering the bursty magnetic enhancement. This is substantially different from the situation of the Cartesian simulation, in which both $[y\Rightarrow\hspace{-4.5mm}_{_x}~\, y]$ and $[y\Rightarrow\hspace{-4.5mm}_{_z}~\, y]$ are always negative, namely the compressible terms work as the decay of $B_{y}^2$ by divergent expansion. 

After $t\gtrsim 83$ rotations, the total compressible effect is negative as the increased magnetic pressure countervails the compressible amplification. After that, the shearing terms take on the role of the magnetic amplification to the 
maximum peak at $t\approx 91$ rotations.

\subsection{Termination of magnetic activity}

\begin{figure*}
   \begin{center}
     \includegraphics[width=11cm]{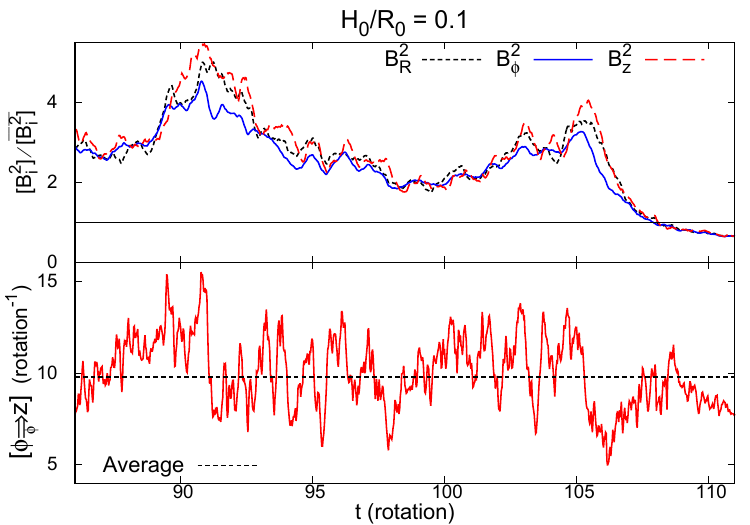}
   \end{center}
   \caption{Time evolution of magnetic energy (top) and the variation rate of $B_z^2$ by shearing from $\phi$ to $z$ (bottom) averaged over 
     the domain in the declining phase of the same active period 
     shown in Figure \ref{fig:dBp2_B2}. What are shown in the top panel are the three components of $[B_{i}^2]$ with $i=R$ (black dotted), $\phi$ (blue solid), and $z$ (red dashed), normalized by the respective time-averaged values, $\overline{[B_{i}^2]}$, during $t=25 - 300$ rotations. $[B_i^2]/\overline[B_i^2] = 1$ is plotted by the black thin solid line. In the bottom panel, the time average, 
     $\overline{[\phi\Rightarrow\hspace{-4mm}_{_{\phi}}~\, z]} = +9.71$
     is also plotted by the black dotted line.
     \label{fig:dBz2_B2}
   }
\end{figure*}

Next, we focus on the declining phase of the same active phase discussed in Section \ref{sec:onset}.
The top panel of Figure \ref{fig:dBz2_B2} presents the three components of magnetic energy averaged over the same narrow region with $R=0.95R_0-1.05R_0$, adopted for Figure \ref{fig:dBp2_B2}. In order to see relative enhancements, each $[B_i^2]$ is normalized by the time-averaged $\overline{[B_i^2]}$.
One can recognize that $[B_z^2]/\overline{[B_z^2]}$ (red dashed line) reaches the highest peaks at $t\approx 90.8$ and $105.4$ rotations among the three components just before the drops of the magnetic energy. The times for these $z$-component peaks occur slightly later than those for the $R$ and $\phi$ components. Moreover, in the subsequent declining period, $[B_z^2]/\overline{[B_z^2]}$ is larger than $[B_R^2]/\overline{[B_R^2]}$ (black dotted line) and $[B_{\phi}^2]/\overline{[B_{\phi}^2]}$ (blue solid line).
In light of these properties on the time evolution, we speculate that the magnetically active states are turned off when the magnetic energy being exchanged between the $R$ and $\phi$ components rapidly leaks to the $z$ component.

In order to verify this speculation, we inspect the transfer rate, 
$[\phi\Rightarrow\hspace{-4mm}_{_{\phi}}~\, z]$, from $B_{\phi}^2$ to $B_{z}^2$ in the bottom panel of Figure \ref{fig:dBz2_B2}. It monotonically increases in the rising phase of $t\le 90$ rotations as $[B_{\phi}^2]$ increases, and it stays at a high level, which 
reaches more than 1.5 times of the time-averaged rate. 
In the descending phase of $90\lesssim t\lesssim 95$ rotations, although the transfer rate is reduced as the energy source, $B_{\phi}^2$, has already declined, it
is maintained at a level comparable to the time-averaged rate, which further reduces $[B_{\phi}^2]$. A qualitatively similar tendency is seen before the later peak at $t\approx 105$ rotations. 
After this time, the transfer rate drops to a low level because the energy source, $B_{\phi}^2$ has sharply decreased earlier than $B_z^2$. The drop in $[\phi\Rightarrow\hspace{-4mm}_{_{\phi}}~\, z]$ reduces $B_z^2$, and the high magnetic activity phase is finally terminated. 
We can conclude that the enhanced leakage of the magnetic energy on the $R-\phi$ plane to the $z$ component 
is a trigger for the end of the high magnetic activity.

\section{Discussion}
\label{sec:dis}

\subsection{Radial dependence of epicyclic frequency}
\label{sec:rdepkapp}
\begin{figure}
  \begin{center}
    \includegraphics[width=9cm]{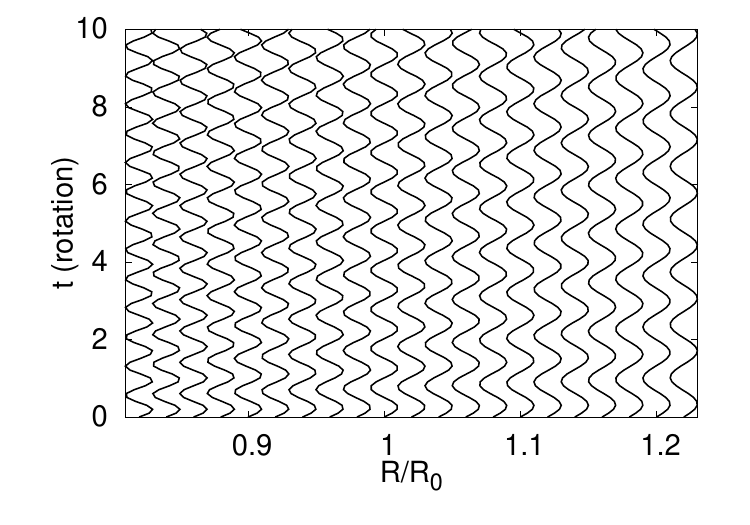}
  \end{center}
  \caption{Radial distance (horizontal) vs. time (vertical) diagram for radial displacements by epicyclic oscillations for $H_0/R_0=0.1$. 
    \label{fig:epc_Rdep01.eps}
  }
\end{figure}

\begin{figure*}
  \begin{center}
    \includegraphics[width=15cm]{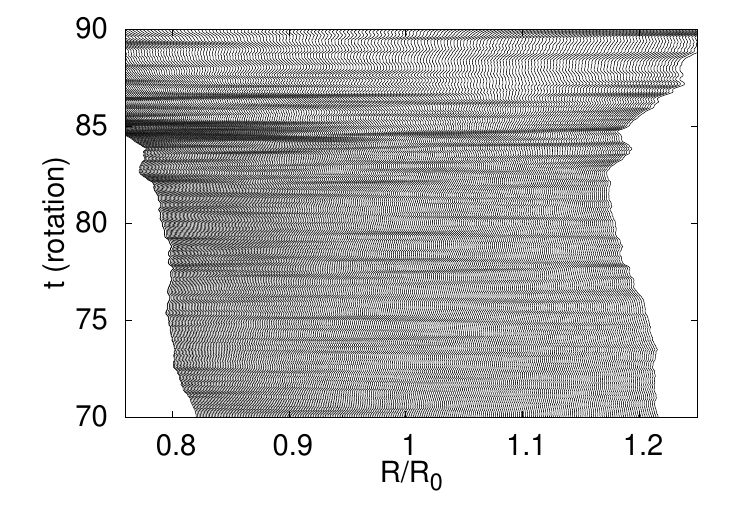}
  \end{center}
  \caption{Radial distance (horizontal) vs. time (vertical) diagram for radial displacements taken from the numerical data of the cylindrical case with $H_0/R_0=0.1$. 
    \label{fig:vR1D}
  }
\end{figure*}

A significant difference of the cylindrical approach from the Cartesian one is the presence of the radial variation in epicyclic frequency. For the equilibrium rotation profile, equation (\ref{eq:Omgeq}), the epicyclic frequency is written as 
\begin{equation}
  \kappa = \sqrt{R\frac{\partial\Omega^2}{\partial R} + 4\Omega^2} = \Omega_{\rm eq} = \Omega_{\rm K}\sqrt{1-2f_{\rm K}}, 
\end{equation}
where we used $f_{\rm K}= H^2/R^2=H_0^2/R_0^2=$ const. (equation \ref{eq:fK}), which is practically satisfied in our simulations.
While in the Cartesian box obviously $\kappa$ is spatially constant, in the cylindrical box $\kappa$ is radially dependent, $\kappa \propto R^{-3/2}$. 

Figure \ref{fig:epc_Rdep01.eps} demonstrates radial displacements, $\xi_R=\xi_0 \sin(\kappa t)$, arising from epicyclic oscillations in the case with $H_0/R_0 = 0.1$, giving $\Omega_{\rm eq}= 0.99\Omega_{\rm K}$, where an arbitrary amplitude is set to be $\xi_0=0.01R_0$.
The figure clearly illustrates that the displacements in neighboring radial locations gradually become out of phase with time because $\frac{\partial \kappa}{\partial R}\ne 0$. Consequently, the phase mixing induces converging and diverging regions by the radial component of the oscillations, which inevitably contributes to the compressible amplification and diffusion of magnetic fields. 

  Figure \ref{fig:vR1D} presents the radial displacements of fluid in the cylindrical case with $H_0/R_0=0.1$ in a time-distance diagram between 70 and 90 rotations. Here the radial displacement of each radial cell is calculated from the $\phi$ and $z$ averaged (equation \ref{eq:pzave}) and mass-weighted $v_R$, 
  \begin{equation}
    \xi_R = \int dt \frac{\langle\rho v_R\rangle}{\langle\rho\rangle}, 
  \end{equation}
  where $\xi_R=0$ is set at the initial time, $t=70$ rotations, of the presented period.  In the inactive period of  $t\lesssim 83$ rotations, the gas accretes inward in a quasi-steady manner. In contrast, when the magnetic activity is enhanced after $t\gtrsim 83$ rotations, the outer gas moves outward while the inner gas continues to accrete; the gas in the domain radially expands, which will be discussed in Section \ref{sec:avsia}. 

  Compared with the ideal setting of ordered epicyclic oscillations in Figure \ref{fig:epc_Rdep01.eps}, radial oscillations in the numerical simulation (Figure \ref{fig:vR1D}) are more randomly excited by turbulence in a stochastic fashion. Accordingly, the picture of the phase mixing of initially ordered epicyclic oscillations based on Figure \ref{fig:epc_Rdep01.eps} has to be generalized. Indeed, one can recognize randomly and ubiquitously formed concentrations of $\xi_R$ trajectories, which are a characteristic feature of converging regions, as seen in the ideal setup (Figure \ref{fig:epc_Rdep01.eps}).  Furthermore, these converging regions are more frequently seen in the elevating phase of the magnetic activity during $t=80-90$ rotation time, which is consistent with the argument on the importance of the compressible amplification in triggering the high magnetic activity (Figure \ref{fig:dBp2_B2} and Section \ref{sec:onset}).

\begin{figure}
  \begin{center}
    \includegraphics[width=9cm]{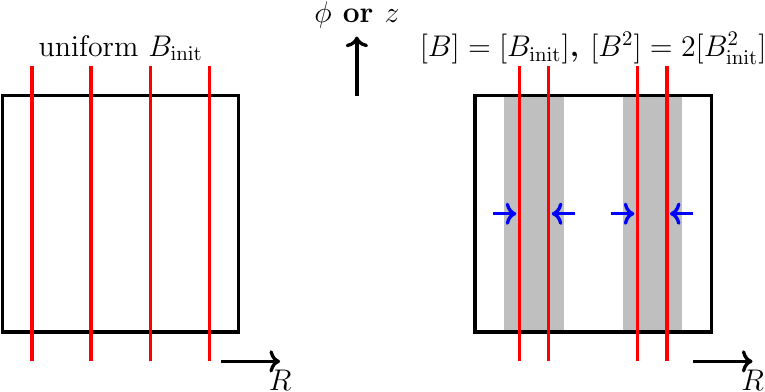}
  \end{center}
  \caption{Conceptual cartoon for the role of the compressible term in the amplification of $B_{\phi}$ (or $B_z$) (red lines). The initial uniform field (left picture) is amplified in converging flows (shaded regions of right picture) and diluted in diverging flows (white regions of right picture) by the radial component of epicyclic oscillations. The magnetic energy $\propto [B_{\phi}^2]$ in the right picture is double of the initial value, while the magnetic field strength $= [B_{\phi}]$ is conserved (see text).  
    \label{fig:Bvconv01}
  }
\end{figure}
Let us suppose a simple situation in which toroidal (or vertical) magnetic field with constant strength, $B_{\phi,{\rm init}}$ (or $B_{z,{\rm init}}$), is initially distributed (left of Figure \ref{fig:Bvconv01}). If we pick out the radial compression term from equation (\ref{eq:Bengp}), the corresponding term of the induction equation is 
\begin{equation}
  \frac{\partial B_{\phi}}{\partial t} = \cdots -\frac{\partial}{\partial R} (v_R B_{\phi}) + \cdots.
\end{equation}
This is essentially the continuity equation for $B_{\phi}$, 
whereas the geometrical curvature term, $\frac{1}{R}\frac{\partial}{\partial R}(R\cdots)$, is omitted if it is compared to the mass continuity equation (\ref{eq:mass}). Therefore, $B_{\phi}$ is compressed (rarefied) in the converging (diverging) regions (right of Figure \ref{fig:Bvconv01}). 
If we compare the left and right pictures of Figure \ref{fig:Bvconv01}, the volume integrated field strength is conserved because the number of the field lines does not change. On the other hand, the magnetic energy becomes double as the regions with strong field, $2B_{\phi,{\rm init}}$, occupy 50\% of the volume (the shaded regions in the right picture of Figure \ref{fig:Bvconv01}) so that $[B_{\phi}^2] = [(2B_{\phi,{\rm init}})^2]\times 0.5 = 2[B_{\phi,{\rm init}}^2]$.

The consideration based on this simple model indicates that the radial compression term should systematically increase $B_{\phi}^2$, which does not occur in the Cartesian setup with the spatially constant $\kappa$.
Although the evolution of the magnetic field is more complex in realistic situations as discussed in Section \ref{sec:res}, this simple mechanism is expected to work in the nonlinear saturated states, and therefore, the radial compression, $\overline{[\phi\Rightarrow\hspace{-4mm}_{_{_R}}~\, \phi]}$, takes the relatively large positive values in the cylindrical simulations, which is in clear contrast to the negative $\overline{ [y\Rightarrow\hspace{-4mm}_{_{x}}~\, y]}$ in the Cartesian simulation (Figure \ref{fig:Triangle01}).
The same argument can be applied to the vertical magnetic field. Indeed, Figure \ref{fig:Triangle01} shows the radial compression of $B_z$, $\overline{[z\Rightarrow\hspace{-4mm}_{_{_R}}~\, z]}$, is positive in the cylindrical cases, which is also in contrast to the negative corresponding term,  $\overline{ [z\Rightarrow\hspace{-4mm}_{_{x}}~\, z]}$, in the Cartesian case. 

The abovementioned argument indicates that the presence of $\frac{\partial \kappa}{\partial R}\ne 0$ significantly changes the fundamental properties of the amplification of magnetic fields. This is the reason why the nearly Cartesian simulation with $H_0/R_0=0.01$ gives the very different conversion rates of $[i\Rightarrow\hspace{-4.mm}_{_k}~\, j]$ from the exact Cartesian case but the similar values to the moderately cylindrical case with $H_0/R_0 = 0.1$ as shown in the triangle diagrams of Figure \ref{fig:Triangle01}. 

The next question one might ask would be regarding the timescale for the onset of  the compressible amplification.
When the oscillatory phases are deviated between radially ``neighboring" regions each other by $\pi$, the compressible amplification works most effectively as shown in Figure \ref{fig:Bvconv01}. We can estimate the time, $t=\tau_{\rm comp}$, to reach the phase difference of $\Delta\phi_{\kappa} = \pi$ from the uniform oscillation at $t=0$ via $\Delta \kappa_0 t = \Delta\phi_{\kappa}$, where the deviation of $\kappa_0$ is derived from $\Delta\kappa = |\left(\frac{\partial \kappa}{\partial R}\right)_0|\Delta R_0$ by using the radial spacing between the neighboring regions, $\Delta R_0$. Since $|\left(\frac{\partial \kappa}{\partial R}\right)_0| = \frac{3\Omega_{\rm eq,0}}{2R}$, we have 
\begin{align}
  \tau_{\rm comp} &= \frac{20\pi}{3} \left(\frac{H_0/R_0}{0.1}\right)^{-1} \left(\frac{\Delta \phi_{\kappa}}{\pi}\right)\left(\frac{\Delta R}{H_0}\right)^{-1} \nonumber \\
  &=\frac{10}{3} \left(\frac{H_0/R_0}{0.1}\right)^{-1} \left(\frac{\Delta \phi_{\kappa}}{\pi}\right)\left(\frac{\Delta R}{H_0}\right)^{-1} [{\rm rotations}]
  \label{eq:t_shift}
\end{align}
where the radial spacing is normalized by $H_0$ for a typical scale of the system. 
This equation shows that the epicyclic oscillation becomes completely out of phase ($\Delta \phi_{\kappa}=\pi$) at radial spacing of $\Delta R=H_0$ when $\tau_{\rm comp}\approx 3$ rotations for the cylindrical case with $H_0/R_0=0.1$.
For $H_0/R_0=0.01$, longer $\tau_{\rm comp} \approx 30$ rotations are required.
However, we would like to note that the estimate based on equation (\ref{eq:t_shift}) is rather conservative because the effect of the radial compression is regarded to be already effective well before the phase difference reaches $\pi$. Additionally, oscillations are expected to be excited rather randomly by turbulence, as discussed with Figure \ref{fig:vR1D}.
Thus, even in this nearly Cartesian case, the compressible amplification is already working at $t\approx 10$ rotation time, which is comparable to the timescale for the transition from the initial linear stage to the nonlinear saturated phase (Figure \ref{fig:alphaM_tevol}).

Another characteristic feature expected from equation (\ref{eq:t_shift}) is that, as time goes on, the compressible amplification is going to work for smaller $\Delta R$; smaller-scale localized magnetic concentrations can be formed at later times. 
In order to capture these fine-scale structures, numerical simulations require sufficiently high resolution. Thus, the saturation level and time-variability of magnetic field may depend on numerical resolution, which will be addressed in our future work.

Global simulations should automatically take the effect of $\frac{\partial \kappa}{\partial R}\ne 0$ into account. Therefore, if one applied the same analyses on the magnetic evolution in Section \ref{sec:evolB} to global simulation data, the similar result regarding the importance of the compressible effect should be obtained, although such an attempt has not been tried within our knowledge. However, here, numerical resolution would matter as discussed above. If the numerical resolution in the radial direction is insufficient and the difference in $\kappa$ between neighboring cells is too large, it appears that intermittent variability cannot be captured (Akatsuka \& Suzuki 2023, in preparation).

\subsection{Substructures and viscous-type instability}
\label{sec:substructure}
\begin{figure*}
  \begin{center}
    \includegraphics[width=15cm]{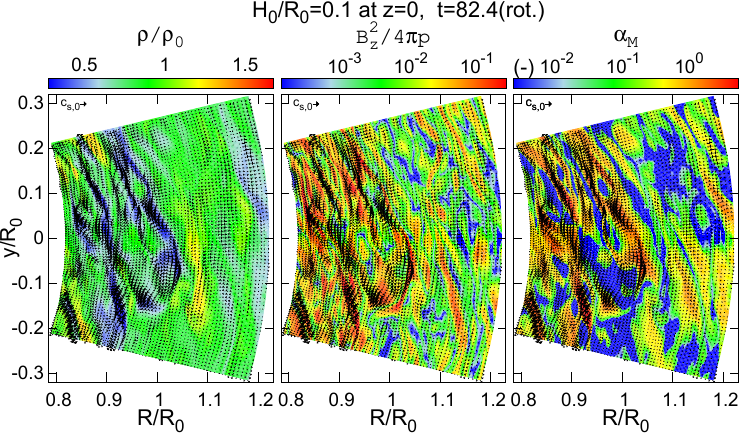}
  \end{center}
  \caption{Face-on view of dimensionless density ($\rho/\rho_0$; left), magnetization with respect of $B_z$ ($B_z^2/4\pi p$; middle), and dimensionless Maxwell stress ($\alpha_{\rm M}$; right) on the midplane ($z=0$) at $t=82.4$ rotations.  Velocity fields are also shown by arrows. We note that the velocity vectors are directed along the perturbation component, $(\delta v_R,\delta v_{\phi}) = (v_R,v_{\phi}-R\Omega_{\rm eq})$ (equation \ref{eq:dvphi}). Movie during $t=70-110$ rotations is available at \url{https://ea.c.u-tokyo.ac.jp/astro/Members/stakeru/research/movie/index.html}.
    \label{fig:velfld}
  }
\end{figure*}

\begin{figure}
  \begin{center}
    \includegraphics[height=6cm]{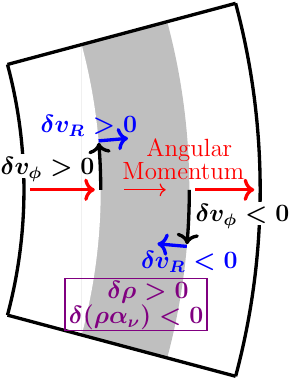}
  \end{center}
  \caption{Physical picture of the viscous-type instability. See text for the detailed explanation.
    \label{fig:visinstb}
  }
\end{figure}

\begin{figure*}
  \begin{center}
    \includegraphics[width=15cm]{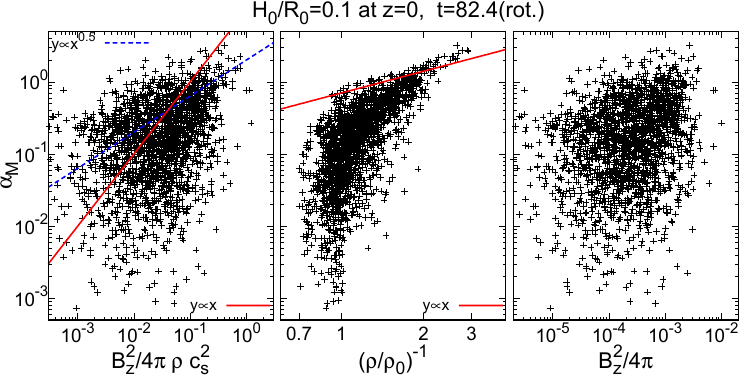} 
  \end{center}
  \caption{Scatter plots of dimensionless Maxwell stress, $\alpha_{\rm M}$, against $B_z^2/4\pi \rho c_{\rm s}^{2}$ (left), $(\rho/\rho_0)^{-1}$ (middle), and $B_z^2/4\pi$ (right) at $t=82.4$ rotations. The data are 
    averaged over 2 radial and 16 azimuthal grid cells of the 2D plots in Figure \ref{fig:velfld}. The slope of $y\propto x$ is overlaid in the left and middle panels by the red solid line, and $y\propto x^{0.5}$ is in the left panel by the blue dashed line.
    \label{fig:B-alp-rho_corr}
  }
\end{figure*}

Figure \ref{fig:Bvconv01} implies that converging and diverging regions move with time. 
In the nonlinear stage, oscillatory motions are excited randomly by turbulence and those at different radial positions undergo phase shift each other. Hence, it is expected that the locations of converging regions change with time in a more or less stochastic manner. 
These randomly excited compressed regions could be seeds of substructures by connecting to various types of instability \citep[e.g.,][see \citet{Lesur2022} for recent review]{Suriano2019MNRAS,Cui2021}. 

Figure \ref{fig:velfld} shows a two-dimensional (2D) $r-\phi$ slice of density, $\rho/\rho_0$ (left), ``magnetization'', $B_z^2/4\pi p$ (middle), defined as the inverse of $\beta_z$ (c.f., equation \ref{eq:betazinit}), and Maxwell stress, $\alpha_{\rm M}$ (right) on $z=0$ at $t=82.4$ rotation time, which is in the onset phase of the high magnetic activity analyzed in Section \ref{sec:onset} (see Figures \ref{fig:alphaM_tevol} \& \ref{fig:dBp2_B2}). 
One can recognize stripe structures with anti-correlated density and magnetic field; low-density and strong-magnetic field regions are sandwiched between denser and weaker-field regions with typical spacing of a fraction of $H_0$. These rings and gaps are ubiquitously and continuously formed as shown in the movie file. Such ring-and-gap structures 
are already seen in early global unstratified ideal MHD simulations \citep{Steinacker2002ApJ} and discussed from a viewpoint of viscous-type instability \citep{Lightman1974} generalized by including the physical properties of the MRI \citep{Hawley2001}.

One of the key ingredients of the viscous-type instability in the MHD framework is magnetic diffusion. Although the resistivity is in principle zero under the ideal MHD approximation, numerical simulations inevitably undergo numerical diffusion. From a physical point of view, reconnections of turbulent magnetic fields would also provide effective magnetic diffusion and dissipation even when the ideal MHD condition is satisfied \citep{Lazarian1999ApJ, Lazarian2020Phpl}. 
We carry out linear perturbation analyses for viscous-type instability with explicitly taking into account resistivity in the induction equation (\ref{eq:ind}). We basically follow the formulation introduced by \citet{Riols2019}, but restrict our analyses to the case without the mass loss and angular momentum removal by MHD disk winds. An important modification from the original setting in \citet{Riols2019} is that dimensionless viscosity, $\alpha_{\nu}$, is assumed to depend separately on density and magnetic field: 
\begin{equation}
  \alpha_{\nu}\propto \left(\frac{\rho_0}{\rho}\right)^{q_{\rho}}\left(\frac{B_z^2}{B_{z,0}^2}\right)^{q_B},
  \label{eq:alphaMdep}
\end{equation}
where we can practically assume $\alpha_{\nu}\approx \alpha_{\rm M}$ to interpret our simulation results.
We apply plane-wave expansions to the linearized equations of mass continuity (equation \ref{eq:mass}), radial and angular momenta (equation \ref{eq:mom}), and vertical magnetic field (equation \ref{eq:ind}) with resistivity. We finally obtain the criterion for the presence of an unstable mode is 
\begin{equation}
  q_{\rho} > 1,
  \label{eq:qrho}
\end{equation}
where the detailed derivation is presented in Appendix \ref{sec:lpa}.

We note that in \citet{Riols2019}  $\alpha_{\nu}$ is assumed to depend on magnetization, $B_z^2/4\pi \rho c_{\rm s}^2$, namely $q=q_{B} = q_{\rho}$ is imposed, and they derived the instability condition as $q > 1$. The relation between time- and volume-averaged $\alpha_{\rm M}$ and $B_z^2/4\pi \rho c_{\rm s}^2$ have been extensively examined with Cartesian shearing box simulations \citep{Salvesen2016,Scepi2018} and global simulations \citep{Mishra2020}. There is a rough consensus of $q\approx 0.5-0.7$, which does not satisfy the instability condition in \citet{Riols2019}. 
On the other hand, our analyses in Appendix \ref{sec:lpa} shows that $q_B$ does not qualitatively affect the stability criterion but only quantitatively controls the growth (or decay) rate.

Figure \ref{fig:visinstb} illustrates the physical picture of the instability. Let us suppose a situation in which a higher-density, $\delta \rho > 0$, region (shaded area) is created by random perturbations. If $\delta (\rho \alpha_{\nu}) < 0$ there, the outward transport of the angular momentum is suppressed in this denser ring (red arrows).  As a result, at the inner (outer) edge of the ring the angular momentum increases (decreases) to give $\delta v_{\phi} > (<)$ $0$  (black arrows), which causes the inner (outer) edge to move outward (inward), $\delta v_R > (<)$ $0$ (blue arrows). Hence, the density of the denser ring further increases. 
If we pick out the dependence on density\footnote{In the ideal MHD condition, the dependence on $B_z$ should also be considered here as $\rho$ and $B_z$ behave ``in phase'' each other. However, the inclusion of resistivity breaks this constraint and it is justified to focus only on the dependence on $\rho$; see Appendix \ref{sec:lpa} for the detailed algebra.} in equation (\ref{eq:alphaMdep}) and take the Taylor expansion,  $\delta (\rho \alpha_{\nu}) \approx (1-q_{\rho})\alpha_{\nu,0}\delta \rho$. Therefore, the condition for the instability, $\delta (\rho \alpha_{\nu}) < 0$, corresponds to $q_{\rho}>1$ (equation \ref{eq:qrho}).

Figure \ref{fig:B-alp-rho_corr} shows scatter plots between $\alpha_{\rm M}$ (vertical axis) and various quantities (horizontal axis) of each grid point displayed in Figure \ref{fig:velfld}.
In the left panel, the correlation with magnetization is shown. The dependence is roughly consistent with the slope derived in the abovementioned previous works, whereas the plots are largely scattered, reflecting the scatter in $B_z^2$ (right panel), because the data are not averaged over time or domain. The middle panel, which shows the dependence on the inverse of density, exhibits relatively tighter correlation particularly in the larger-$\alpha_{\rm M}$ and lower-density (upper right) side. Moreover, the slope is slightly steeper than $q_{\rho}=1$ and is in the unstable regime (equation \ref{eq:qrho}). We examined scatter plots for $\rho$-$\alpha_M$ in other time frames and found that $q_{\rho}\gtrsim 1$ is kept in most of the time. Therefore, we can interpret that the substructures seen in our simulations (Figure \ref{fig:velfld}) are amplified with a secular timescale 
(see Appendix \ref{sec:lpa} for the derivation) and maintained in the cylindrical disk that are in the marginally unstable condition regarding the generalized viscous-type instability.

\subsection{Intermittency}
\label{sec:intermittency}
Cartesian shearing box simulations for MRI often show quasi periodicities with $t\sim 10$ rotation time 
\citep{Davis2010ApJ,Gressel2010,Guan2011ApJ}, which is related to the recurrent growth and breakup of large-scale channel-mode flows \citep{Gogichaishvili2018ApJ}. 
While quasi-periodic magnetic activity is more clearly seen in vertically stratified simulations, being frequently associated with mass outflows \citep[e.g.,][]{Suzuki2009ApJ,Suzuki2010ApJ,Wissing2022}, similar periodicity is also observed in unstratified simulations \citep{Sano2001,Shi2016}.  Our Cartesian case is also showing mild time variation in $[\alpha_{\rm M}]$ (black dashed line in Figure \ref{fig:alphaM_tevol}). 

On the other hand, the duration $\sim 50$ rotations between high-magnetic-activity phases seen in the cylindrical case with $H_0/R_0=0.1$ (red solid line in Figure \ref{fig:alphaM_tevol}) is considerably longer than the typical period $\sim$ 10 rotations described above. 
A speculative mechanism for this weak and long-term periodicity is related to the compressible effect arising from the radial variation of $\kappa$ (Section \ref{sec:rdepkapp}). If we utilize equation (\ref{eq:t_shift}), the time to reach the phase difference, $\Delta \phi_{\kappa}=2\pi$, between neighboring radial regions can be estimated as a possible source of the periodicity. If we adopt $\Delta R\approx 0.2 R_0$, referring to the radial width of a ring (or a gas) in Figure \ref{fig:velfld}, we obtain $t\approx 33$ rotation time, which is roughly consistent with the observed cycle of the enhanced magnetic activitye. On the other hand, this estimate based on the initially ordered epicyclic oscillations (Figure \ref{fig:epc_Rdep01.eps} Section \ref{sec:rdepkapp}) may not be relevant, because oscillations driven constantly and randomly by turbulence interact each other as illustrated in Figure \ref{fig:vR1D}.

\subsection{Accretion structure in active and inactive phases}
\label{sec:avsia}
\begin{figure*}
  \includegraphics[width=9cm]{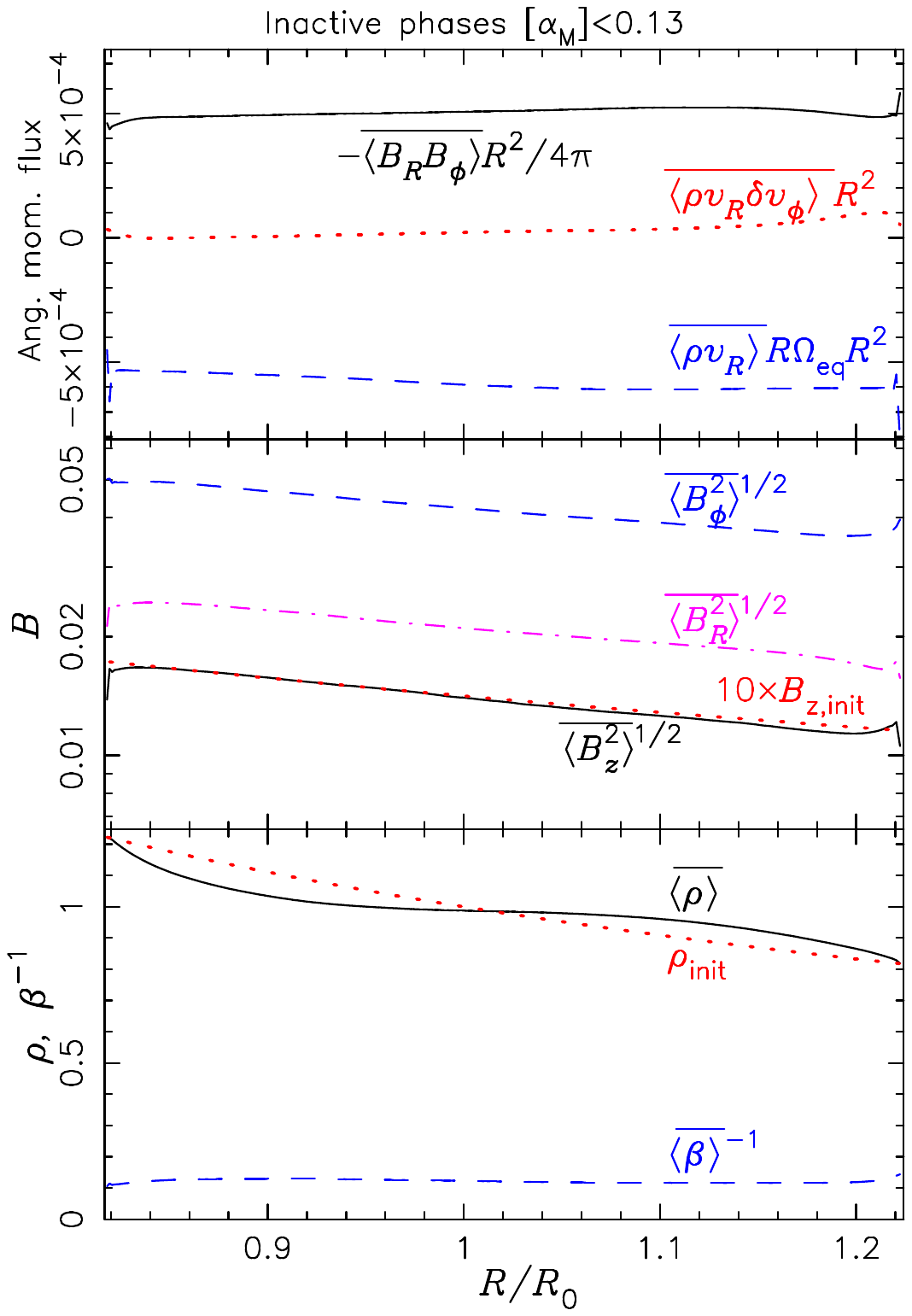}
  \includegraphics[width=9cm]{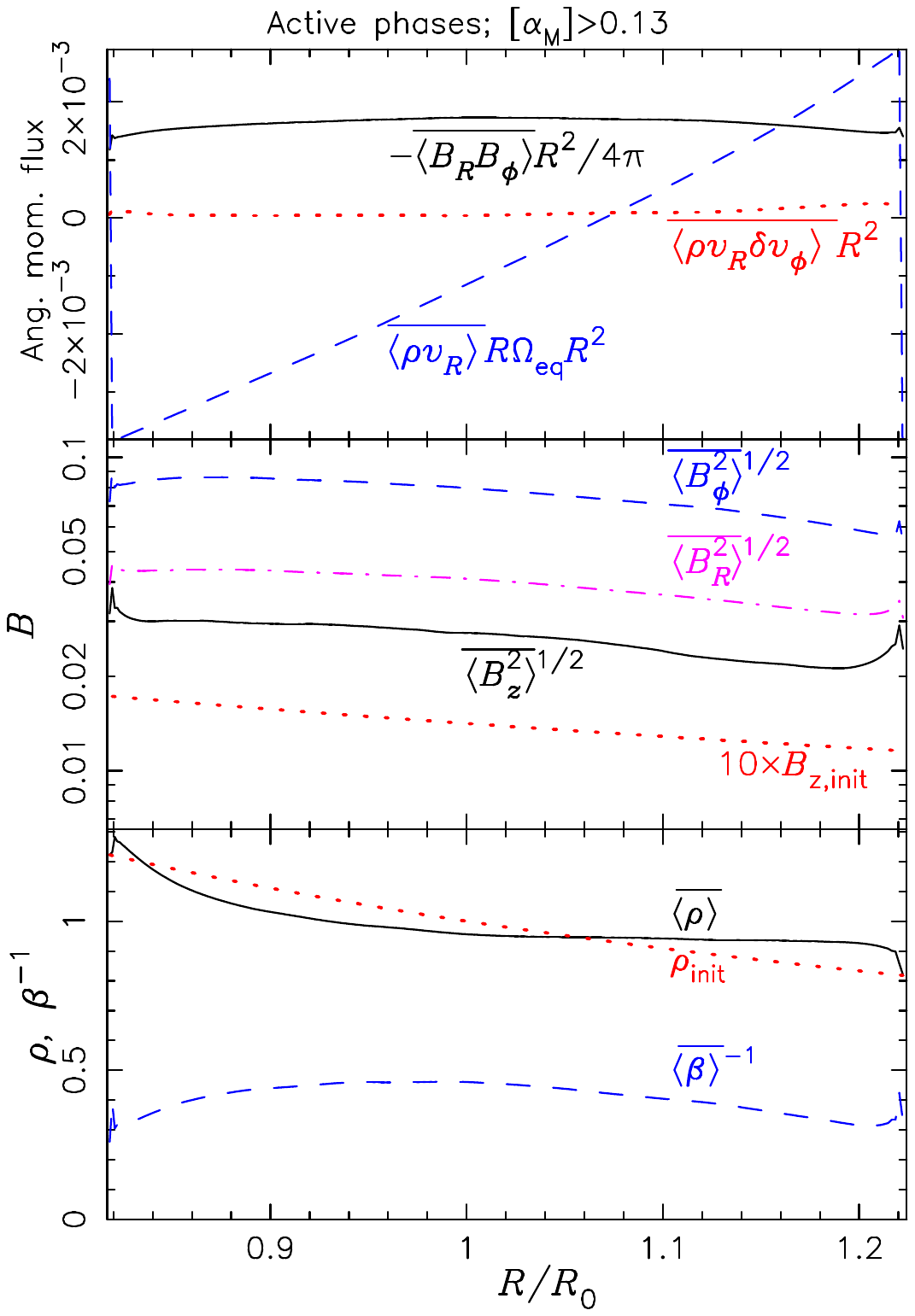}
  \caption{Comparison of time-averaged radial profiles of various physical quantities of the case with $H_0/R_0=0.1$ in the inactive (left) and active (right) periods.
    The top panels show angular momentum fluxes carried by Maxwell stress (black solid), Reynolds stress (red dotted), and mean accretion flow (blue dashed).
    The middle panels compare rms $B_{R}$ (magenta dash-dotted), $B_{\phi}$ (blue dashed), and $B_z$ (black solid). The initial vertical field strength, $B_{z,{\rm init}}$, is multiplied by a factor of 10 (red dotted), in order to compare with the field strength in the saturated state.
    The bottom panels show density (black solid) in comparison to the initial distribution (red dotted) and the inverse of plasma $\beta = 8\pi p/B^2$ (blue dashed). 
    We note that the vertical scales of the top and middle panels are different for the left and right panels.   See Section \ref{sec:basics} for the physical units on the vertical axes.
    \label{fig:rad1D_ina_act}
  }
\end{figure*}

\begin{figure*}
  \begin{center}
    \includegraphics[height=7cm]{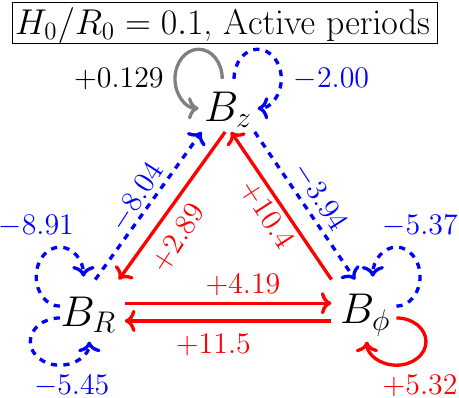}
    \includegraphics[height=7cm]{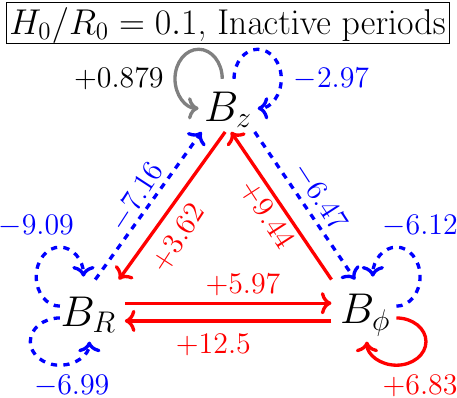}\\
  \end{center}
  \caption{Triangle diagrams for the cylindrical case with $H_0/R_0 = 0.1$ (the top-right triangle in Figure \ref{fig:Triangle01}). The left and right triangles are the results of the active ($[\alpha_{\rm M}]>0.13$) and and inactive ($[\alpha_{\rm M}]<0.13$) periods.
\label{fig:Triangle02}
  }
\end{figure*}

We compare the radial dependences of various physical quantities during magnetically active and inactive phases in Figure \ref{fig:rad1D_ina_act}. In order to see the difference between the active and inactive periods, we take the temporal averages for $[\alpha_{\rm M}]< 0.13$ (left) and $[\alpha_{\rm M}]> 0.13$ (right). The inactive (active) phase occupies 246.2 (28.8) rotations out of the total time-averaged duration of $275(=300-25)$ rotations.

The top panels of Figure \ref{fig:rad1D_ina_act} compare the radial profiles of angular momentum fluxes, following S19. When the time-steady condition is satisfied, an equation for angular momentum integrated over the $\phi$-$z$ plane is
\begin{equation}
  \frac{\partial}{\partial R}\left[R^2\left(\overline{\langle \rho v_R\rangle}R\Omega_{\rm eq}
    + \overline{\langle \rho v_R\delta v_{\phi}\rangle} - \frac{\overline{\langle B_R B_{\phi}\rangle}}{4\pi}\right)\right] = 0, 
  \label{eq:angmomblnc}
\end{equation}
where the first (blue dashed), second (red dotted), and third (black solid) terms indicate the angular momentum carried by the mean accretion flow, the turbulent Reynolds stress, and the Maxwell stress, respectively.
As shown in the top left panel, the steady accretion structure is well realized in the inactive phase; the inward angular momentum flux by the accretion flow is balanced with the outward flux by the Maxwell stress with a small contribution from the Reynolds stress.

On the other hand, in the active phase (top right panel) the steady accretion structure appears to be broken. This is mainly because the increased magnetic pressure pushes gas inward (outward) from the inner (outer) boundary as discussed below. However, the averaged radial velocity, $\overline{\langle v_R \rangle}$, near the inner and outer boundaries is only a few \% of the sound velocity\footnote{In deriving this value, we note that $\rho c_s R^3\Omega_{\rm eq}\approx 0.1$ in the simulation units for $c_{\rm s,0}/R\Omega_{\rm K,0}=H_0/R_0 = 0.1$ (Section \ref{sec:model})}. This gradual expanding velocity is much smaller than the fluctuating velocity, which is nearly an order of the sound speed in the active phase. 
 
The middle panels of Figure \ref{fig:rad1D_ina_act} show each component of the rms magnetic field strength, $\sqrt{\overline{\langle B_i^{2}\rangle}}$, in comparison to the initial $B_{z,{\rm init}}$. The left panel indicates that the vertical magnetic field strength, $\sqrt{\overline{\langle B_z^{2}\rangle}}$, averaged over the inactive periods increases about 10 times  from the initial value with almost keeping the initial profile, $\propto R^{-1}$. The radial and azimuthal field strengths are larger (Table \ref{tab:Bvalues}) but the radial dependences are similar to that of the $z$ component.
In the active phase (middle right panel), although the field strength is larger as expected, 
the ratios between different components is not so different from those in the inactive phases. 
The toroidal component exhibits a weakly concave down profile. Then, the direction of the magnetic pressure gradient is inward near the inner boundary, while it is directed outward in the intermediate and outer regions. These cause the slow expanding flows seen in the top right panel.

In the bottom panel of Figure \ref{fig:rad1D_ina_act}, density and the inverse of plasma $\beta$, 
\begin{equation}
\overline{\langle\beta\rangle}^{-1} = \frac{\overline{\langle B^2\rangle}}{8\pi \overline{\langle\rho\rangle}}, 
\end{equation}
are presented. The density structure is moderately altered from the initial condition, $R^{-1}$, as it is affected by temporal zonal flows; the density in the inner region slightly decreases, while it increases in the outer region. As a result, $\overline{\langle\beta\rangle}^{-1}$ exhibits a weak radial dependence, which can be more easily seen in the active phase. The radial distribution of the gas pressure is also modified according to the altered density profile. This also leads to the alternation of the time-averaged rotation profile from $\Omega_{\rm eq}$, which affects the estimate of the Reynolds stress in equation \ref{eq:Reynolds} as discussed in Section \ref{sec:tevol}.

We present triangle diagrams for the active and inactive periods in Figure \ref{fig:Triangle02}. One can find that these two phases exhibit quite similar tendencies. Careful comparison shows that most of the variation rates, $\frac{\partial \ln [B_i^2]}{\partial t}$, are slightly smaller in the active phase. This is because these numerical values are normalized by larger $[B_i^2]$; if we compare the change of the magnetic energies, $\frac{\partial [B_i^2]}{\partial t}$, the active phase yields larger values. 

\subsection{Boundary effect}
\label{sec:bdef}
There is a freedom to set the fluctuation amplitudes at the radial boundaries (Section \ref{sec:rdshbd}). We determined the parameters to control these boundary amplitudes in order that the steady accretion structure is reproduced when we take the time average over the inactive periods (right panels of Figure \ref{fig:rad1D_ina_act}), which is described in detail in Appendix \ref{sec:modbd}. 
On the other hand, as discussed above, the steady accretion is not achieved for the radial structure averaged over the active periods. This raises doubts whether the obtained results may be affected by the boundary treatment.  Hence, we perform cylindrical simulations with $H_0/R_0=0.1$ for different values of the amplitude parameters in Appendix \ref{sec:modbd}.

Although these different cases show different radial flow structures and yield stochastically enhanced magnetic activities at different times, we obtain similar properties of the intermittency; magnetic bursts occasionally appear in low-activity phases that occupy most of the simulation time. Consequently, these different cases give very similar time averaged Maxwell stresses and magnetic field strengths (Figures \ref{fig:alpha_3cmp} \& \ref{fig:rad1D_3cmp} and Table \ref{tab:B4famp} in Appendix \ref{sec:modbd}). Furthermore, the obtained variation rates of the magnetic energies, $[i\Rightarrow\hspace{-4.mm}_{_k}~\, j]$, in the triangle diagrams are also very similar each other (Figure \ref{fig:Triangle_3cmp}). 

So far, we have examined the $[i\Rightarrow\hspace{-4.mm}_{_k}~\, j]$ terms integrated over the whole simulation domain. However, these terms may be spatially dependent particularly near the the radial boundaries. 
  To inspect the boundary effect, we compare all $[i\Rightarrow\hspace{-4.mm}_{_k}~\, j]$ terms of the case with $H_0/R_0=0.1$ ($0.01$) in a smaller radial region between $R=0.90R_0$ ($0.990R_0$) and $1.11R_0$ ($1.010R_0$) 
  to those averaged over the entire simulation domain shown in Figure \ref{fig:Triangle01}.
  Two winding terms, $\overline{[R\Rightarrow\hspace{-4mm}_{_{_R}}~\, \phi]}$ and $\overline{[z\Rightarrow\hspace{-4mm}_{_z}~\, \phi]}$, in the case with $H_0/R_0=0.1$ are slightly affected; specifically, the averages in the narrow region are $\overline{[R\Rightarrow\hspace{-4mm}_{_{_R}}~\, \phi]} = 4.12$ and $\overline{[z\Rightarrow\hspace{-4mm}_{_z}~\, \phi]}=-4.41$, which are both reduced by $\approx 20 \%$ from the domain-averaged values. On the other hand, the modifications in the other winding terms and all the compressible terms are less than 5\%. In the case with $H_0/R_0=0.01$, the modifications of all the terms are less than 5\%, because the magnetic activity is relatively weak so that the contribution from the active periods causing the mismatched radial boundaries is almost ignorable. In summary, we can conclude that the effect regarding the boundary treatment is limited and then the discussion based on the numerical results so far is unaffected.

\section{Summary}
\label{sec:sum}
Continuing from S19, 
we studied fundamental MHD properties of accretion disks by cylindrical shearing box simulations.
We modified the treatment for the radial boundary condition from the original prescription in S19. The key improvement is to separate the boundary variables into the mean and perturbation components and the amplitude of the latter is adjusted to match the fluctuations at both boundaries (Section \ref{sec:rdshbd} and Appendix \ref{sec:modbd}). This modified prescription enables us to reduce unmatched fluctuations traveling across the boundaries.

The radial gradient of epicyclic frequency, $\kappa$, causes the phase mixing of random oscillatory motions. As a result, the radial compression is significant in amplifying the azimuthal and vertical magnetic fields (Section \ref{sec:rdepkapp}). This is an important finding in this paper by inspecting the spatial derivative terms on the right-hand-side of the equations for magnetic energy. 
In contrast, the compressible effect works as diverging dilution of the magnetic energy in the Cartesian box simulation because of the absence of the radial variation of $\kappa$ (Section \ref{sec:evolB}).  

The compressible amplification plays a significant role in enhanced bursty magnetic activity, which is more clearly seen in the case with larger $H_0/R_0$ (Section \ref{sec:tevol}). This is expected from the argument on the timescale of the phase shift due to 
the non-uniform distribution of $\kappa$ (Section \ref{sec:rdepkapp}). We also speculate that the phase-shift timescale is related to the weak periodicity in the intermittent magnetic bursts (Section \ref{sec:intermittency}).

The compressible amplification is also expected to create seeds of small-scale substructures. Indeed, there are narrow ring-gap structures continuously and ubiquitously formed in the simulations. These structures show the anti-correlation between density and magnetic field strength; in particular, the steep dependence of Maxwell stress on density, $\alpha_{\rm M} \propto \rho^{-q_\rho}$, with $q_{\rho}\gtrsim 1$ is obtained (Section \ref{sec:substructure}).
We revisited the viscous-type instability by considering the dependence of $\alpha_{\rm M}$ separately on density and vertical magnetic field, and found that the instability condition is $q_{\rho}>1$ (Appendix \ref{sec:lpa}). 
Thus, we interpreted that the ring-gap structures are maintained in the simulated disks that are under marginally unstable conditions.

In this paper, as we focused on the effects of the disk curvature, $H_0/R_0$, we did not conduct simulations with different initial vertical field strengths, box sizes, or numerical resolutions. However, the dependences on these parameters are obviously important, which will be addressed in our future papers.

A key physics in our work is the radial gradient of $\kappa$; its nonlinear outcomes are time-variability, intermittency, and localized substructures observed in the simulations. This problem can be framed as nonlinear processes in a system where the eigen-mode frequency varies spatially. When the physical quantities are non-uniformly distributed, as being found in various astrophysical systems, such as interstellar medium in star-forming regions and the interior, atmosphere and magnetosphere of stars and compact objects, eigen-mode frequencies of various oscillatory modes are also expected to be spatially dependent. This study could have potential applications in such systems, which is one of the future directions.


The author thanks the anonymous referee for valuable comments to the original draft. 
Numerical computations were carried out on Cray XC50 at Center for Computational Astrophysics, National Astronomical Observatory of Japan, and Yukawa-21 (Dell PowerEdge R840) at YITP, Kyoto University.
This work is supported by Grants-in-Aid for Scientific Research from the MEXT/JSPS of Japan, 17H01105, 21H00033 and 22H01263 and by Program for Promoting Research on the Supercomputer Fugaku by the RIKEN Center for Computational Science (Toward a unified view of the universe: from large-scale structures to planets; grant 20351188-PI J. Makino) from the MEXT of Japan.

\appendix
\section{Modified shearing radial boundary condition}
\label{sec:modbd}
\begin{figure}
  \begin{center}
    \includegraphics[height=6cm]{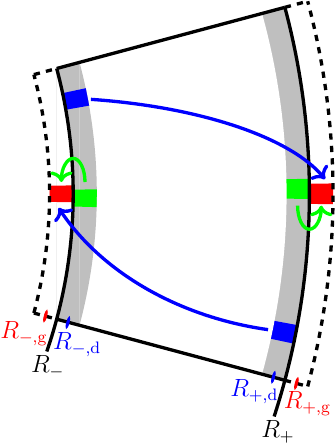}
  \end{center}
  \caption{Shearing boundary condition in a local cylindrical box. The simulation domain, which is enclosed by the solid black line, is between $R=R_-$ and $R_+$. $R_{-,{\rm d}}$ ($R_{+,{\rm d}}$) is the radial location of the innermost (outermost) cells in the simulation domain, which are illustrated by gray shade. $R_{-,{\rm g}}$ ($R_{+,{\rm g}}$) is the radial location of the inner (outer) ghost cells. The physical quantities of the ghost cell at $R=R_{\pm,{\rm g}}$ (red) are determined from those of the corresponding sheared cells at $R=R_{\mp,{\rm d}}$ (blue) and the radially adjacent cell(s) at $R=R_{\pm,{\rm d}}$ (green). See text for the detail.
    \label{fig:cylbox}
  }
\end{figure}

We describe the specific method to implement the shearing boundary condition, equation (\ref{eq:rdbdsum}).
As we stated in Section \ref{sec:rdshbd}, the modification from the original prescription in S19 is that the shearing variables are separated into the mean values and perturbations (equation \ref{eq:prtb}).
We rewrite equation (\ref{eq:rdbdsum}) in a more precise way to distinguish ghost (``g'') and domain (``d'') cells (Figure \ref{fig:cylbox}):  
\begin{align}
  S\mbox{\footnotesize ($R_{\pm,{\rm g}},\phi$)} &= \langle S\mbox{\footnotesize ($R_{\mp,{\rm d}},\phi  - (\Omega_{\rm eq,\pm,g} - \Omega_{\rm eq,\mp,d})t$)}\rangle + \delta S \nonumber\\
  &\approx \langle S\mbox{\footnotesize ($R_{\mp,{\rm d}},\phi  + \Delta \Omega_{\rm eq}t$)}\rangle + \delta S, 
\end{align}
where we omitted ``$z$'' in the arguments as it is redundant and can be understood without it.
We employ the standard periodic boundary condition for the mean components, $\langle S\rangle$: the shearing variables averaged over the innermost (outermost) $\phi$-$z$ plane inside the simulation domain (gray shaded regions in Figure \ref{fig:cylbox}) are passed to the outer (inner) ghost cells outside the domain (regions surrounded by the dashed lines in Figure \ref{fig:cylbox}).

On the other hand, for the perturbation components, $\delta S$, in a ghost cell at $R_{\pm,{\rm g}}$(red in Figure \ref{fig:cylbox}), we consider the contribution from not only the corresponding sheared cells at $R_{\mp,{\rm d}}$ (blue) but also the adjacent cells at $R_{\pm,{\rm d}}$ (green).
Specifically, we adopt the following parameterization: 
\begin{align}
  \delta S\mbox{\footnotesize ($R_{\pm,{\rm g}},\phi$)} &= f_{\rm amp,\pm}\left[f_{\rm sh}K\delta S\mbox{\footnotesize ($R_{\mp,{\rm d}},\phi  - (\Omega_{\rm eq,\pm,g} - \Omega_{\rm eq,\mp,d})t$)}  + (1-f_{\rm sh})\delta S\mbox{\footnotesize ($R_{\pm,{\rm d}},\phi -(\Omega_{\rm eq,\pm,g} - \Omega_{\rm eq,\pm,d})t$)}\right] \label{eq:dSexact} \\
  &\approx f_{\rm amp,\pm}\left[f_{\rm sh}K\delta S\mbox{\footnotesize ($R_{\mp,{\rm d}},\phi + \Delta\Omega_{\rm eq} t$)}  + (1-f_{\rm sh})\delta S\mbox{\footnotesize ($R_{\pm,{\rm d}},\phi$)}\right] \label{eq:dSapprox}, 
\end{align}
where $f_{\rm amp,\pm}(\approx 1)$ is a parameter that controls the amplitude of $\delta S$,  $f_{\rm sh}$ determines the fractional contributions from the sheared position at the opposite side of the $R$ domain (blue in Figure \ref{fig:cylbox}), and 
\begin{equation}
  K = \sqrt\frac{\langle\delta S^2\mbox{\footnotesize ($R_{\pm,{\rm d}},\phi -(\Omega_{\rm eq,\pm,g} - \Omega_{\rm eq,\pm,d})t$)} \rangle}{\langle\delta S^2\mbox{\footnotesize ($R_{\mp,{\rm d}},\phi  - (\Omega_{\rm eq,\pm,g} - \Omega_{\rm eq,\mp,d})t$)}\rangle} \approx  \sqrt\frac{\langle\delta S^2\mbox{\footnotesize ($R_{\pm,{\rm d}},\phi$)} \rangle}{\langle\delta S^2\mbox{\footnotesize ($R_{\mp,{\rm d}},\phi  \pm \Delta\Omega_{\rm eq}t$)}\rangle}.
  \label{eq:Kcorr}
\end{equation}
Since in general the relative azimuthal position between the ghost and domain cells, which changes with time, is not an exact integer multiple of the azimuthal size of the fixed grid cell, $\delta S$'s in the first and second terms on the right-hand side of equation (\ref{eq:dSexact}) need to be interpolated from two grid cells. However, as we ignore the tiny difference between $\Omega_{\rm eq,\pm,d}$ and $\Omega_{\rm eq,\pm,g}$ and use equation (\ref{eq:dSapprox}) in the numerical implementation,  $\delta S$ in the second term is taken from the adjacent grid cell in the $R$ direction.

We note that $f_{\rm sh}=1$ recovers the original treatment of S19, and $f_{\rm sh}=0$ corresponds to the ``non-gradient'' boundary condition for $\delta S$. Throughout the current paper, we adopt $f_{\rm sh}=0.5$, namely the fluctuations at both the shearing location and the neighboring position are equally mixed.   
We also note that, when $f_{\rm amp,\pm}=1$, the correction by this $K$ ensures that the rms amplitude averaged over the ghost cells on $R=R_{\pm,{\rm g}}$ matches that averaged over the neighboring cells at $R=R_{\pm,{\rm d}}$:  
$\sqrt{\delta S^2\mbox{\footnotesize ($R_{\pm,{\rm g}},\phi$)}} = \sqrt{\delta S^2\mbox{\footnotesize ($R_{\pm,{\rm d}},\phi$)}}$.
MRI grows from earlier times at smaller $R$. Without this amplitude correction, it can be easily seen that fluctuations in the inner regions leak out of $R=R_{-}$ and seeps from $R=R_+$ to the quiet outer region in the initial growth phase, as exhibited in Figure 1 of S19.
\begin{figure}
  \begin{center}
    \includegraphics[height=6cm]{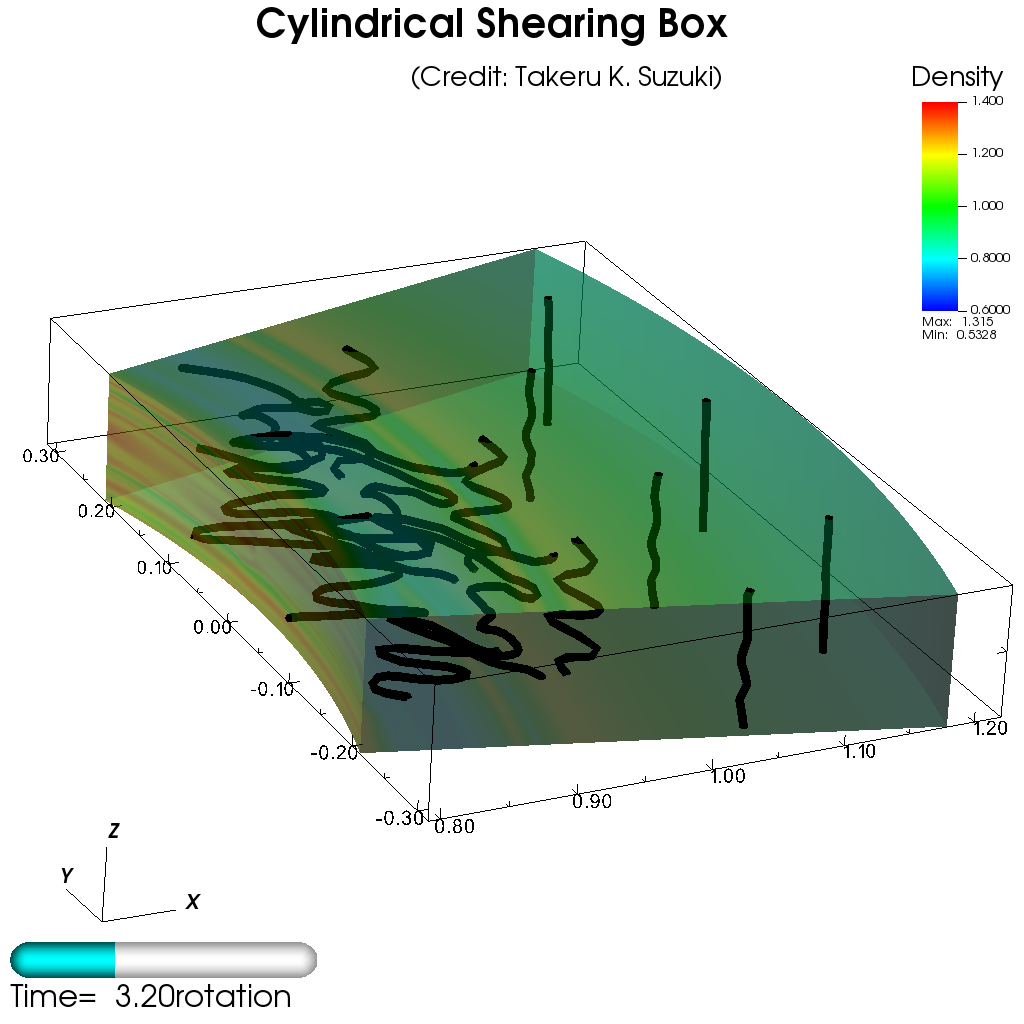}
  \end{center}
  \caption{3D snapshot of the case with $\beta_{z.{\rm init}}=10^3$ and the vertical box size, $L_z=H$, at $t=3.20$ rotation. 
    {\it Movie} between 0 and 10 rotations is available at \url{https://ea.c.u-tokyo.ac.jp/astro/Members/stakeru/research/movie/index.html}.
    \label{fig:cylbdtest}
  }
\end{figure}
Figure \ref{fig:cylbdtest} shows the initial growing stage of MRI with the amplitude correction by equation (\ref{eq:Kcorr}). The simulation setup is the same as in S19: $\beta_{z,{\rm init}}=10^3$ and the vertical domain size, $L_z=H$, is smaller than that adopt in the present paper.  The figure is demonstrating that the contamination of the inner fluctuations passed to the outer boundary, $R=R_+$, is greatly suppressed, compared with that found in S19\footnote{As the FARGO advection scheme adopted in this paper suffers from less numerical diffusion than the normal advection scheme used in S19, MRI sets in from slightly earlier time. Thus, we are showing the snapshot at $t=3.20$ rotations, which is earlier than $t=3.45$ and $4.00$ rotations presented in S19.}. 

\begin{figure*}
  \begin{center}
    \includegraphics[height=6.6cm]{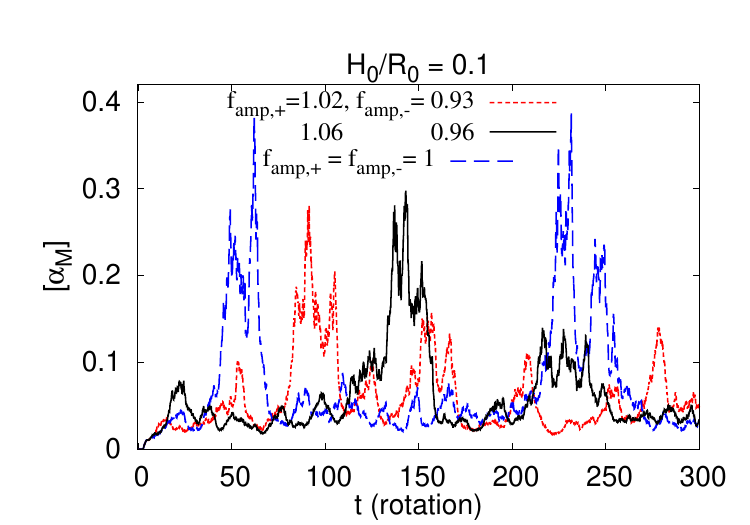}
  \end{center}
  \caption{Same as Figure \ref{fig:alphaM_tevol} but for cases with different $f_{\rm amp,\pm}$ for $H_0/R_0=0.1$. The red dotted, black solid, and blue dashed lines are for $(f_{\rm amp,+}, f_{\rm amp,-})=(1.02,0.93)$, $(1.06,0.96)$, and (1,1), respectively. Note that the first case (red dotted) corresponds to the same one presented in the main text (the red solid line in figure \ref{fig:alphaM_tevol}).
\label{fig:alpha_3cmp}
  }
\end{figure*}

\begin{figure*}
  \includegraphics[width=6.6cm]{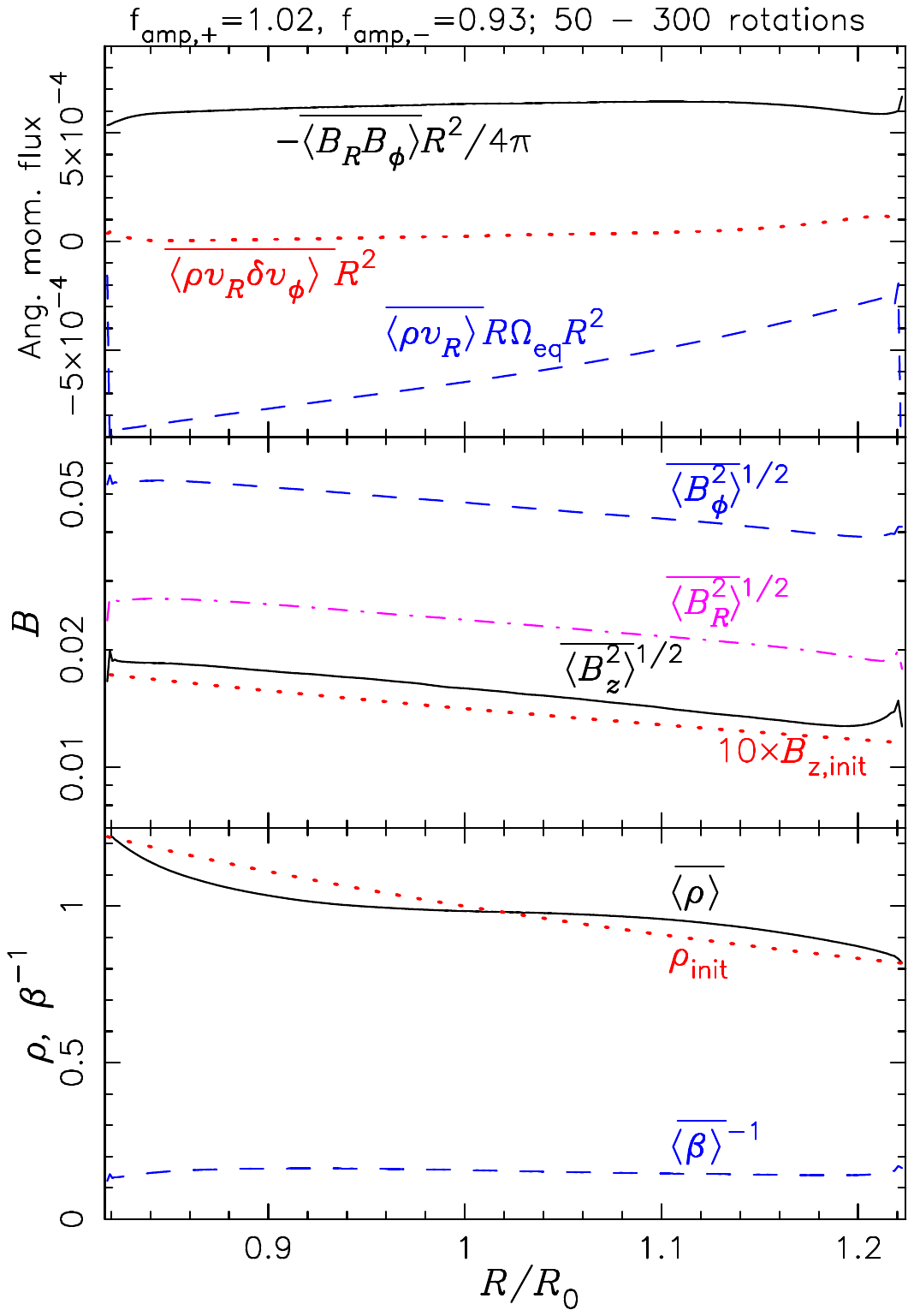}
  \hspace{-1cm}
  \includegraphics[width=6.6cm]{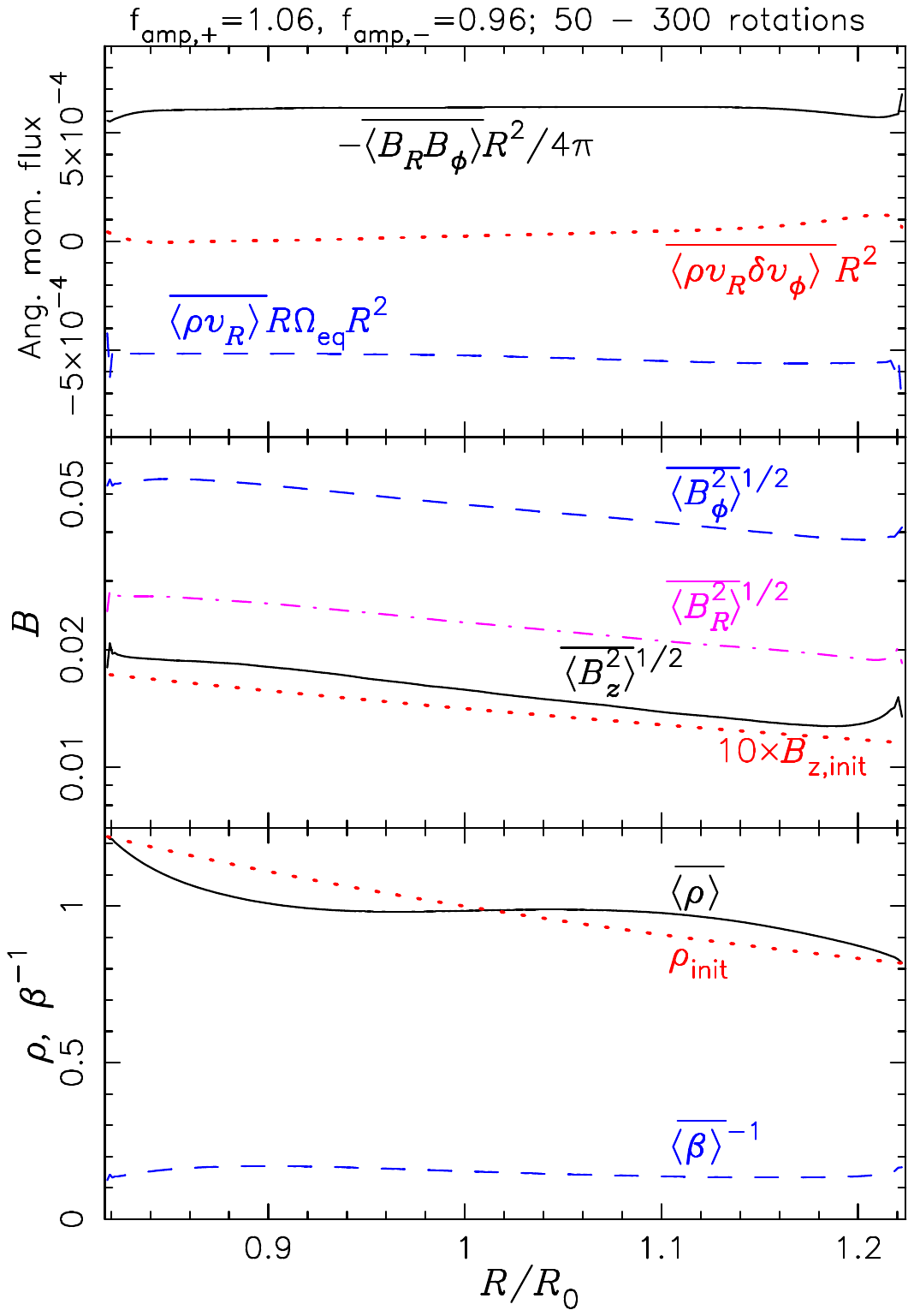}
  \hspace{-1cm}
  \includegraphics[width=6.6cm]{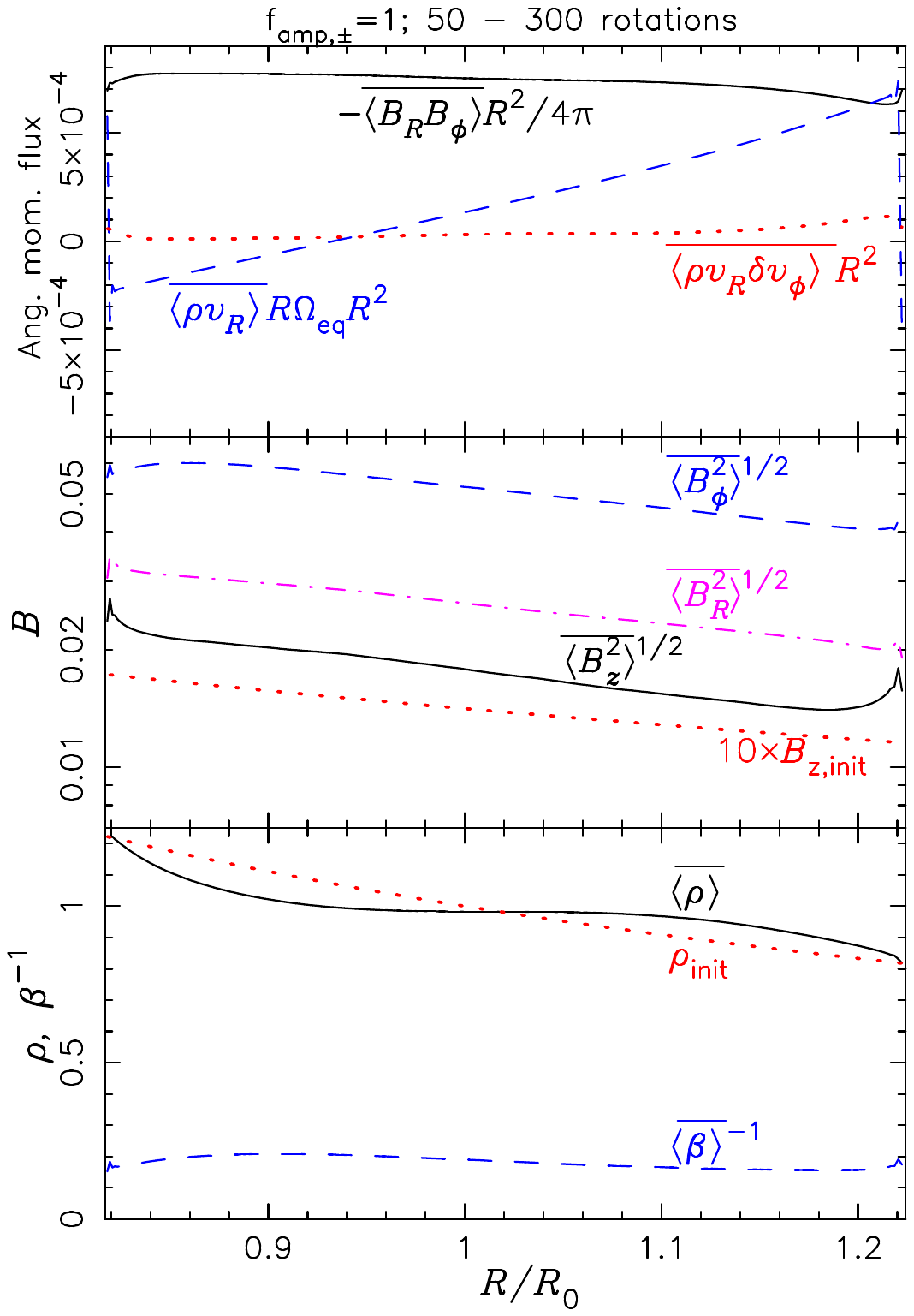}
  \caption{Same as Figure \ref{fig:rad1D_ina_act} but for cases with different $f_{\rm amp,\pm}$ averaged over 25 -- 300 rotations. The left, middle, and right panels are for $(f_{\rm amp,+}, f_{\rm amp,-})=(1.02,0.93)$, $(1.06,0.96)$, and (1,1), respectively.
    Note that the physical quantities in the left panels are the average of those in both active and inactive phases shown in Figure \ref{fig:rad1D_ina_act}. 
    \label{fig:rad1D_3cmp}
  }
\end{figure*}

\begin{figure*}
  \begin{center}
    \includegraphics[width=5.4cm]{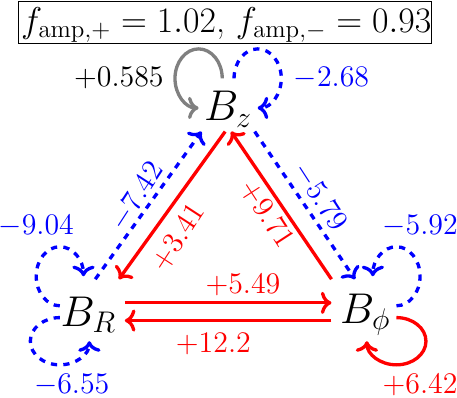}
    \includegraphics[width=5.4cm]{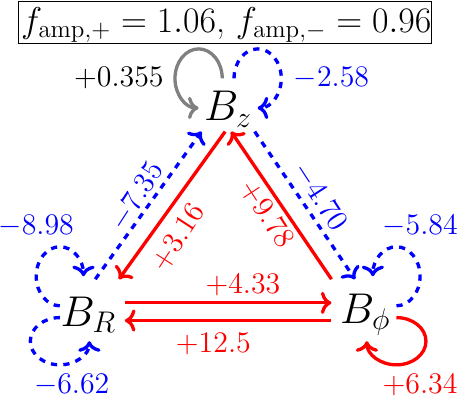}
    \includegraphics[width=5.4cm]{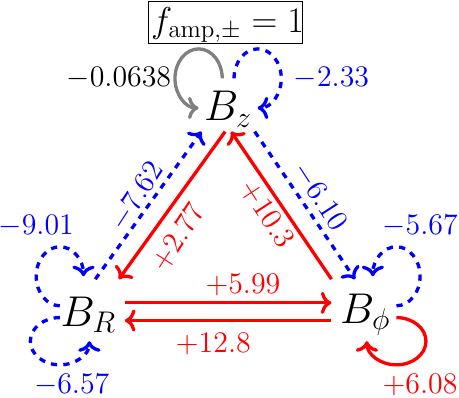}
  \end{center}
  \caption{Same as Figure  \ref{fig:Triangle01} but for cases with different $f_{\rm amp,\pm}$ for $H_0/R_0=0.1$. The left, middle, and right triangles correspond to those in Figure \ref{fig:rad1D_3cmp}. Note that the left triangle is the same as the top right triangle in  Figure \ref{fig:rad1D_ina_act}. 
\label{fig:Triangle_3cmp}
  }
\end{figure*}

\begin{table*}
  \begin{center}
    \begin{tabular}{cccccccc}
      \hline
      $(f_{\rm amp,+},f_{\rm amp,-})$ & $\overline{[\alpha_{\rm M}]}$  & $\overline{[B_R^2]} / 8\pi\overline{[ p ]}$ & $\overline{[B_{\phi}^2]} / 8\pi\overline{[ p ]}$ & $\overline{[B_z^2]} / 8\pi\overline{[ p ]}$ & $\overline{[B_R^2]} / \overline{[B_{\phi}^2]}$ & $\overline{[B_z^2]} / \overline{[B_{\phi}^2]}$ & $\overline{[\alpha_{\rm R}]}$\\
      \hline
      \hline
      $(1.02,093)$ & $6.22\times 10^{-2}$ & $2.78 \times 10^{-2}$ & $1.11\times 10^{-1}$ & $1.25\times 10^{-2}$ & $0.250$ & $0.112$ & $3.36\times 10^{-3}$ \\
      $(1.06,0.96)$ & $6.13\times 10^{-2}$ & $2.74\times 10^{-2}$ & $1.09\times 10^{-1}$ & $1.25\times 10^{-2}$ & $0.250$ & $0.114$ & $3.34\times 10^{-3}$\\
      $(1,1)$ & $7.43\times 10^{-2} $ & $3.41\times 10^{-2}$ & $1.32\times 10^{-1}$ & $1.59\times 10^{-2}$ & $0.259$ & $0.120$ & $3.63\times 10^{-3}$\\
      \hline
    \end{tabular}
    \caption{Same as Table \ref{tab:Bvalues} but for cases with different $f_{\rm amp,\pm}$ for $H_0/R_0=0.1$. Note that the case with $(f_{\rm amp,+}, f_{\rm amp,-})=(1.02,0.93)$ is the same as the first case presented in Table \ref{tab:Bvalues}. 
    \label{tab:B4famp}
    }
  \end{center}
\end{table*}

$f_{\rm amp,\pm}$ is tuned to realize the steady accretion structure as shown in the top left panel of Figure \ref{fig:rad1D_ina_act}. Larger $f_{\rm amp,\pm}$ yields larger perturbations in the ghost zone at $R=R_{\pm,{\rm g}}$, which results in larger effective turbulent pressure there. Therefore, larger $f_{\rm amp,+}$ ($f_{\rm amp,-}$) tends to induce inward (outward) flows.  We are adopting $f_{\rm amp,+}=1.02$ and $f_{\rm amp,-}=0.93$ for the cylindrical case with $H_0/R_0=0.1$ and $f_{\rm amp,+}=0.94$ and $f_{\rm amp,-}=0.93$ for $H_0/R_0=0.01$.

We demonstrate how different choices of $f_{\rm amp,\pm}$ affect the results of the case with $H_0/R_0=0.1$ in Figures \ref{fig:alpha_3cmp} -- \ref{fig:Triangle_3cmp} and Table \ref{tab:B4famp}.  We perform simulations with $(f_{\rm amp,+},f_{\rm amp,-}) = (1.06,0.96)$ and $(1,1)$ in addition to the original case with $(f_{\rm amp,+},f_{\rm amp,-})=(1.02,0.93)$.
  From Figure \ref{fig:alpha_3cmp} we find that, although the timings of magnetic enhancements are different, these three cases exhibit similar intermittent properties.

The top panels of Figure \ref{fig:rad1D_3cmp} indicates that the radial accretion profile is affected by the values of $f_{\rm amp,\pm}$. In the original case (left), although the steady accretion profile is achieved in the inactive periods, the radial flow shows a gentle expanding trend by the contribution from the active periods (Figure \ref{fig:rad1D_ina_act}). On the other hand,  the case with $(f_{\rm amp,+},f_{\rm amp,-}) = (1.06,0.96)$ yields a steady accretion feature\footnote{In this case, the temporal average during the inactive periods gives a gentle converging trend, which cancels a slow expanding profile obtained from the active periods.}, because the expanding flow is confined by a small increase in both $f_{\rm amp,+}$ and $f_{\rm amp,-}$ as discussed above. At the same time, one can also find from the middle and bottom panels of Figure  \ref{fig:rad1D_3cmp} and Table \ref{tab:B4famp} that these two cases give very similar time-averaged magnetic fields.

The case with $f_{\rm amp,\pm}=1$ (right panels of Figure  \ref{fig:rad1D_3cmp}) exhibits an expanding trend with positive $v_R$ in the outer region, which is expected from the smaller $f_{\rm amp,+}$ and larger $f_{\rm amp,-}$ than those in the original case. The magnetic properties of this case are similar to those of the other two cases, although the time-averaged field strength is slightly larger (Table \ref{tab:B4famp}) by the larger contribution of the high magnetic activity periods (Figure \ref{fig:alpha_3cmp}).

The triangle diagrams of these three cases show very similar shearing and compressible amplification rates (Figure \ref{tab:B4famp}); in particular, the critical importance of the radial compression in the amplification of $B_{\phi}$ is universally attained.
In summary, although $f_{\rm amp, \pm}$ controls the accretion profile, the intermittency of the magnetic activity and the time-averaged field strength 
are not significantly affected,  provided that $f_{\rm amp,\pm}\approx 1$ is employed. 

\section{Linear perturbation analyses for viscous-type instability}
\label{sec:lpa}

We conduct linear perturbation analyses on the 2D $r-\phi$ plane under the axisymmetric approximation. In the momentum equation (\ref{eq:mom}), we do not explicitly include magnetic fields but consider them through $\alpha$ prescription \citep{Shakura73}. 
Then, the radial and angular momentum equations can be written as 
\begin{equation}
  \frac{\partial}{\partial t}(\rho w_R) = 2\rho \Omega_{\rm K} w_{\phi} -\frac{\partial}{\partial R}(\rho c_{\rm s}^2) +\rho \frac{w_{\phi}^2}{R}
  \label{eq:rmom}
\end{equation}
and
\begin{equation}
  \frac{\partial}{\partial t}(\rho w_{\phi}R) + \frac{1}{R}\frac{\partial}{\partial R}(R^2\rho c_{\rm s}^2\alpha_{\nu}) + \frac{1}{2}\rho\Omega_{\rm K}w_R R = 0 
\label{eq:phimom}
\end{equation}
in Eulerian forms, where $(w_R,w_{\phi}) = (v_R, v_{\phi}-R\Omega_{\rm K})$ is the velocity deviation from Keplerian rotation and $\alpha_{\nu}$ is dimensionless viscosity, equation (\ref{eq:alphaMdep}). We note that, in equation (\ref{eq:rmom}), the inward gravity, $-\rho\frac{GM}{R^2}$, and the centrifugal force, $\rho R \Omega_{\rm K}^2$, should be present but they are exactly canceled out each other.
We adopt the radial profiles of $c_{\rm s}^2\propto R^{-1}$ (equation \ref{eq:Tprfl}), $\rho\propto R^{-1}$, (equation \ref{eq:rhoinit}) and $\alpha_{\nu}\propto R^{1/2}$ (S19) to match our simulation setting. 
Then, from the unperturbed state ($\partial_t = 0$) of equations (\ref{eq:rmom}) and (\ref{eq:phimom}), we obtain
\begin{equation}
w_{\phi} = R\Omega_{\rm K}(\sqrt{1-f_{\rm K}} - 1) \propto R^{-\frac{1}{2}}, 
\end{equation}
which is consistent with equation (\ref{eq:Omgeq}), and
\begin{equation}
  w_R (= v_R) =  -\frac{\alpha_{\nu} c_{\rm s}^2}{R\Omega_{\rm K}}  = -\frac{\alpha_{\nu}H^2\Omega_{\rm K}}{R} \propto R^0.
  \label{eq:w_R}
\end{equation}
$\alpha_{\nu}$ is assumed to depend separately on $\rho$ and $B_z$ (equation \ref{eq:alphaMdep}), and thus, we have 
\begin{equation}
  \alpha_{\nu} \approx \alpha_{\nu,0}  \left(\frac{R}{R_0}\right)^{\frac{1}{2}}\left(1-q_{\rho}\frac{\delta \rho}{\rho_0} + 2q_B\frac{\delta B_z}{B_{z,0}}\right) \equiv \alpha_{\nu,0}^{\prime}\left(1-q_{\rho}\frac{\delta \rho}{\rho_0} + 2q_B\frac{\delta B_z}{B_{z,0}}\right).
  \label{eq:alphanu}
\end{equation}
Here, density is calculated from the continuity equation (\ref{eq:mass}) and vertical magnetic field from the induction equation (\ref{eq:ind}) with resistivity, $\eta$, explicitly included:  
\begin{equation}
\frac{\partial B_z}{\partial t} = \frac{1}{R}\frac{\partial}{\partial R}\left[R\left(-w_R B_z + \eta\frac{\partial B_z}{\partial R}\right)\right], 
\label{eq:indres}
\end{equation}
where we assumed $v_z=0$. We also adopt the $\alpha$ prescription for $\eta$ with the same dependence on $\rho$ and $B_z$,
\begin{equation}
  \eta = \alpha_{\eta}\frac{c_{\rm s}^2}{\Omega_{\rm K}} = \alpha_{\eta}H^2\Omega_{\rm K}= \eta_0\left(\frac{R}{R_0}\right)\left(1-q_{\rho}\frac{\delta \rho}{\rho_0} + 2q_B\frac{\delta B_z}{B_{z,0}}\right).
  \label{eq:eta}
\end{equation}
We note that $\eta\propto R$ ensures that equation (\ref{eq:indres}) has an unperturbed-state solution of $B_z\propto R^{-1}$, which is consistent with equation (\ref{eq:Bzinit}) (see also S19). 

We expand $\rho$, $w_R$, $w_{\phi}$, and $B_z$ with the radial dependences that match our simulation setting: 
\begin{alignat}{2}
  \rho &= (\rho_0 + \delta \rho)\left(\frac{R}{R_0}\right)^{-1} & &\equiv\rho_0^{\prime} + \delta \rho^{\prime},\\
  w_{R} &= w_{R,0} + \delta w_R & &\equiv w_{R,0}^{\prime} + \delta w_R^{\prime},\\
  w_{\phi} &= (w_{\phi,0} + \delta w_{\phi}) \left(\frac{R}{R_0}\right)^{-\frac{1}{2}} & &\equiv w_{\phi,0}^{\prime} + \delta w_{\phi}^{\prime},\\
  B_z &= (B_{z,0} + \delta B_z)\left(\frac{R}{R_0}\right)^{-1} & &\equiv B_{z,0}^{\prime} + \delta B_z^{\prime}.
  \label{eq:dBzexp}
\end{alignat}
We apply these expansions to equations (\ref{eq:mass}), (\ref{eq:rmom}), (\ref{eq:phimom}), and (\ref{eq:indres}) and further apply plane-wave expansion, 
\begin{equation}
  \delta \propto \exp(i\omega t - i kR),
  \label{eq:plwv}
\end{equation}
to the first-order variables. Then, the corresponding four equations are
\begin{equation}
  i(\omega -k w_{R,0}^{\prime})\delta \rho^{\prime} - i k \rho_0^{\prime} \delta w_R^{\prime} =0, 
  \label{eq:masspw}
\end{equation}

\begin{equation}
  i\omega \rho_0^{\prime}\delta w_R^{\prime} - 2\rho_0^{\prime}\left[\Omega_{\rm K} + \frac{w_{\phi,0}^{\prime}}{R}\right]\delta w_{\phi}^{\prime} + \left[i\omega w_{R.0}^{\prime}-2\Omega_{\rm K}w_{\phi,0}^{\prime} - c_{\rm s}^2\left(\frac{2}{R} + ik\right) - \frac{{w_{\phi,0}^{\prime}}^2}{R}\right]\delta\rho^{\prime} = 0,
\end{equation}

\begin{equation}
  i\omega \rho_0^{\prime} \delta w_{\phi}^{\prime} + \frac{1}{2}\Omega_{\rm K}\rho_0^{\prime}\delta w_R^{\prime} + \left[i\omega w_{\phi,0}^{\prime} + \frac{1}{2}\Omega_{\rm K}w_{R,0}^{\prime} + c_{\rm s}^2\alpha_{\nu}^{\prime} (1-q_{\rho})\left(\frac{1}{2R} - ik\right)\right]\delta \rho^{\prime} + c_{\rm s}^2\alpha_{\nu}^{\prime}\frac{\rho_0^{\prime}}{B_{z,0}^{\prime}}\left(\frac{1}{R} - 2ik\right)q_{B}\delta B_z^{\prime} = 0,
\end{equation}
and
\begin{equation}
  \left(i\omega - ikw_{R,0}^{\prime} +\eta_0 k^2 + ik\eta_0 (1-2q_B)\frac{1}{R} \right)\delta B_z^{\prime} - ikB_{z,0}^{\prime}\delta w_R^{\prime} + ik \eta_0 q_{\rho}\frac{1}{R}\frac{B_{z,0}^{\prime}}{\rho_0^{\prime}}\delta \rho^{\prime} =0.
  \label{eq:Bzpw}
\end{equation}
From equations (\ref{eq:masspw}) - (\ref{eq:Bzpw}), we can derive a fourth-order equation with respect to $\omega$.
Here, we ignore the solutions of $\omega^2 \approx \Omega_{\rm K}^2 + k^2 c_{\rm s}^2$ describing sound waves in shearing systems, and focus on secular modes by assuming $\omega^2 \ll \Omega_{\rm K}^2$. In addition, we also assume $H/R \ll 1$ and ignore the terms  $\propto 1/R, 1/R^2, \cdots$ arising from the curvature of cylindrical coordinates.
Then, we finally obtain
\begin{align}
  \omega^2(1+k^2H^2) &- \omega\left[(k w_{R} + ik^2 \eta) (1+k^2 H^2) +2i (1-q_{\rho} + 2q_B) k^2 H^2 \Omega_{\rm K}\alpha_{\nu}\right] \nonumber \\
  &+ 2 \Omega_{\rm K}\alpha_{\nu}k^2H^2 \left[(1-q_{\rho} + 2q_B)ikw_{R} - (1-q_{\rho})k^2\eta\right] = 0,
  \label{eq:wquad}
\end{align}
where we omitted the subscripts, 0, and the superscripts, $\prime$, because we ignored the $R$-dependent terms. 
Substituting equations (\ref{eq:w_R}) and (\ref{eq:eta}) into equation (\ref{eq:wquad}), we can derive
\begin{equation}
\omega = \frac{k^2H^2\Omega_{\rm K}\alpha_{\nu}}{2(1+k^2H^2)}\left[i\left\{2(1-q_{\rho}+2q_B) + \frac{1+k^2H^2}{{\rm Pm}}\right\} \pm \sqrt{-\left\{2(1-q_{\rho}+2q_B) + \frac{1+k^2H^2}{{\rm Pm}}\right\}^2 + 8(1-q_{\rho})\frac{1+k^2H^2}{{\rm Pm}}}\right], 
\label{eq:omegasol}
\end{equation}
where 
\begin{equation}
{\rm Pm} \equiv \frac{\alpha_{\nu}}{\alpha_{\eta}}
\end{equation}
is the magnetic Prandtl number and we once again disregarded the terms $\propto 1/R$ that originally involved $w_R$.
As shown in Figure \ref{fig:B-alp-rho_corr}, $q_{\rho}\approx 1$, and then, in the square-root of equation (\ref{eq:omegasol}) the first term should dominate the second term. Thus, we take the Taylor expansion of the square-root term. Recalling the form of the plane-wave expansion (equation \ref{eq:plwv}), one can find a growing mode, if it is present, only for the negative sign of equation (\ref{eq:omegasol}). The growth rate can be calculated as
\begin{equation}
  s =i\omega = \frac{2(q_{\rho}-1)k^2H^2}{2(1-q_{\rho}+2q_B){\rm Pm}+1+k^2H^2}\Omega_{\rm K}\alpha_{\nu} =
  \begin{cases}
    2(q_{\rho}-1)\frac{k^2H^2}{1+k^2H^2}\Omega_{\rm K}\alpha_{\nu} & {\rm Pm}\ll 1\\
    \frac{(q_{\rho}-1)}{(1-q_{\rho}+2q_B)}k^2H^2\Omega_{\rm K}\alpha_{\eta} & {\rm Pm}\gg 1\\
  \end{cases}
  .
  \label{eq:growthrate}
\end{equation}
This result 
shows that the growth (or decay) rate is determined by the slower process of viscous or resistive diffusion; it is an order of $\Omega_{\rm K}\alpha_{\nu}$ for Pm $\ll 1$ and $\Omega_{\rm K}\alpha_{\eta}$ for Pm $\gg 1$. In the former case, an unstable mode is present if $q_{\rho} > 1$ (equation \ref{eq:qrho}), and the growth rate does not depends on $q_{B}$. In the latter case, although it depends on $q_B$, the instability condition is again $q_{\rho} > 1$ because $(1-q_{\rho}+2q_B)>0$ is probably satisfied in usual situations (Figure \ref{fig:B-alp-rho_corr}).


\bibliography{cylvscar2023}{}

\begin{thebibliography}{}
\expandafter\ifx\csname natexlab\endcsname\relax\def\natexlab#1{#1}\fi
\providecommand{\url}[1]{\href{#1}{#1}}
\providecommand{\dodoi}[1]{doi:~\href{http://doi.org/#1}{\nolinkurl{#1}}}
\providecommand{\doeprint}[1]{\href{http://ascl.net/#1}{\nolinkurl{http://ascl.net/#1}}}
\providecommand{\doarXiv}[1]{\href{https://arxiv.org/abs/#1}{\nolinkurl{https://arxiv.org/abs/#1}}}

\bibitem[{{Adachi} {et~al.}(1976){Adachi}, {Hayashi}, \& {Nakazawa}}]{ada76}
{Adachi}, I., {Hayashi}, C., \& {Nakazawa}, K. 1976, Progress of Theoretical
  Physics, 56, 1756, \dodoi{10.1143/PTP.56.1756}

\bibitem[{{Armitage}(1998)}]{Armitage1998}
{Armitage}, P.~J. 1998, \apjl, 501, L189, \dodoi{10.1086/311463}

\bibitem[{{Bacchini} {et~al.}(2022){Bacchini}, {Arzamasskiy}, {Zhdankin},
  {Werner}, {Begelman}, \& {Uzdensky}}]{Bacchini2022}
{Bacchini}, F., {Arzamasskiy}, L., {Zhdankin}, V., {et~al.} 2022, \apj, 938,
  86, \dodoi{10.3847/1538-4357/ac8a94}

\bibitem[{{Bai} \& {Stone}(2013)}]{Bai2013}
{Bai}, X.-N., \& {Stone}, J.~M. 2013, \apj, 767, 30,
  \dodoi{10.1088/0004-637X/767/1/30}

\bibitem[{{Balbus} \& {Hawley}(1991)}]{Balbus1991}
{Balbus}, S.~A., \& {Hawley}, J.~F. 1991, \apj, 376, 214,
  \dodoi{10.1086/170270}

\bibitem[{{Balbus} \& {Hawley}(1998)}]{Balbus1998}
---. 1998, Reviews of Modern Physics, 70, 1, \dodoi{10.1103/RevModPhys.70.1}

\bibitem[{{Bambic} {et~al.}(2023){Bambic}, {Quataert}, \&
  {Kunz}}]{Bambic2023arxiv}
{Bambic}, C.~J., {Quataert}, E., \& {Kunz}, M.~W. 2023, arXiv e-prints,
  arXiv:2304.06067, \dodoi{10.48550/arXiv.2304.06067}

\bibitem[{{Ben{\'\i}tez-Llambay} \& {Masset}(2016)}]{Benitez2016ApJS}
{Ben{\'\i}tez-Llambay}, P., \& {Masset}, F.~S. 2016, \apjs, 223, 11,
  \dodoi{10.3847/0067-0049/223/1/11}

\bibitem[{{B{\'e}thune} {et~al.}(2017){B{\'e}thune}, {Lesur}, \&
  {Ferreira}}]{Bethune2017}
{B{\'e}thune}, W., {Lesur}, G., \& {Ferreira}, J. 2017, \aap, 600, A75,
  \dodoi{10.1051/0004-6361/201630056}

\bibitem[{{Bodo} {et~al.}(2008){Bodo}, {Mignone}, {Cattaneo}, {Rossi}, \&
  {Ferrari}}]{Bodo2008}
{Bodo}, G., {Mignone}, A., {Cattaneo}, F., {Rossi}, P., \& {Ferrari}, A. 2008,
  \aap, 487, 1, \dodoi{10.1051/0004-6361:200809730}

\bibitem[{{Brandenburg} {et~al.}(1995){Brandenburg}, {Nordlund}, {Stein}, \&
  {Torkelsson}}]{Brandenburg1995ApJ}
{Brandenburg}, A., {Nordlund}, A., {Stein}, R.~F., \& {Torkelsson}, U. 1995,
  \apj, 446, 741, \dodoi{10.1086/175831}

\bibitem[{{Brandenburg} {et~al.}(1996){Brandenburg}, {Nordlund}, {Stein}, \&
  {Torkelsson}}]{Brandenburg96}
---. 1996, \apjl, 458, L45, \dodoi{10.1086/309913}

\bibitem[{{Chandrasekhar}(1961)}]{Chandrasekhar1961}
{Chandrasekhar}, S. 1961, {Hydrodynamic and hydromagnetic stability} (Oxford:
  Clarendon)

\bibitem[{{Clarke}(1996)}]{Clarke96}
{Clarke}, D.~A. 1996, \apj, 457, 291, \dodoi{10.1086/176730}

\bibitem[{{Courant} {et~al.}(1928){Courant}, {Friedrichs}, \& {Lewy}}]{CFL1928}
{Courant}, R., {Friedrichs}, K., \& {Lewy}, H. 1928, Mathematische Annalen,
  100, 32, \dodoi{10.1007/BF01448839}

\bibitem[{{Cui} \& {Bai}(2021)}]{Cui2021}
{Cui}, C., \& {Bai}, X.-N. 2021, \mnras, 507, 1106,
  \dodoi{10.1093/mnras/stab2220}

\bibitem[{{Davis} {et~al.}(2010){Davis}, {Stone}, \& {Pessah}}]{Davis2010ApJ}
{Davis}, S.~W., {Stone}, J.~M., \& {Pessah}, M.~E. 2010, \apj, 713, 52,
  \dodoi{10.1088/0004-637X/713/1/52}

\bibitem[{{Dempsey} {et~al.}(2022){Dempsey}, {Li}, {Mishra}, \&
  {Li}}]{Dempsey2022}
{Dempsey}, A.~M., {Li}, H., {Mishra}, B., \& {Li}, S. 2022, \apj, 940, 155,
  \dodoi{10.3847/1538-4357/ac9d92}

\bibitem[{{Ebrahimi} \& {Blackman}(2016)}]{Ebrahimi2016MNRAS}
{Ebrahimi}, F., \& {Blackman}, E.~G. 2016, \mnras, 459, 1422,
  \dodoi{10.1093/mnras/stw724}

\bibitem[{{Evans} \& {Hawley}(1988)}]{Evans88}
{Evans}, C.~R., \& {Hawley}, J.~F. 1988, \apj, 332, 659, \dodoi{10.1086/166684}

\bibitem[{{Flock} {et~al.}(2013){Flock}, {Fromang}, {Gonz{\'a}lez}, \&
  {Commer{\c{c}}on}}]{Flock2013}
{Flock}, M., {Fromang}, S., {Gonz{\'a}lez}, M., \& {Commer{\c{c}}on}, B. 2013,
  \aap, 560, A43, \dodoi{10.1051/0004-6361/201322451}

\bibitem[{{Fromang} {et~al.}(2013){Fromang}, {Latter}, {Lesur}, \&
  {Ogilvie}}]{Fromang2013}
{Fromang}, S., {Latter}, H., {Lesur}, G., \& {Ogilvie}, G.~I. 2013, \aap, 552,
  A71, \dodoi{10.1051/0004-6361/201220016}

\bibitem[{{Gogichaishvili} {et~al.}(2018){Gogichaishvili}, {Mamatsashvili},
  {Horton}, \& {Chagelishvili}}]{Gogichaishvili2018ApJ}
{Gogichaishvili}, D., {Mamatsashvili}, G., {Horton}, W., \& {Chagelishvili}, G.
  2018, \apj, 866, 134, \dodoi{10.3847/1538-4357/aadbad}

\bibitem[{{Goldreich} \& {Lynden-Bell}(1965)}]{Goldreich1965}
{Goldreich}, P., \& {Lynden-Bell}, D. 1965, \mnras, 130, 125,
  \dodoi{10.1093/mnras/130.2.125}

\bibitem[{{Gressel}(2010)}]{Gressel2010}
{Gressel}, O. 2010, \mnras, 405, 41, \dodoi{10.1111/j.1365-2966.2010.16440.x}

\bibitem[{{Guan} \& {Gammie}(2011)}]{Guan2011ApJ}
{Guan}, X., \& {Gammie}, C.~F. 2011, \apj, 728, 130,
  \dodoi{10.1088/0004-637X/728/2/130}

\bibitem[{{Guilet} {et~al.}(2022){Guilet}, {Reboul-Salze}, {Raynaud}, {Bugli},
  \& {Gallet}}]{Guilet2022}
{Guilet}, J., {Reboul-Salze}, A., {Raynaud}, R., {Bugli}, M., \& {Gallet}, B.
  2022, \mnras, 516, 4346, \dodoi{10.1093/mnras/stac2499}

\bibitem[{{Hawley}(2001)}]{Hawley2001}
{Hawley}, J.~F. 2001, \apj, 554, 534, \dodoi{10.1086/321348}

\bibitem[{{Hawley} {et~al.}(1995){Hawley}, {Gammie}, \& {Balbus}}]{Hawley1995}
{Hawley}, J.~F., {Gammie}, C.~F., \& {Balbus}, S.~A. 1995, \apj, 440, 742,
  \dodoi{10.1086/175311}

\bibitem[{{Hirose} {et~al.}(2009){Hirose}, {Krolik}, \& {Blaes}}]{Hirose2009}
{Hirose}, S., {Krolik}, J.~H., \& {Blaes}, O. 2009, \apj, 691, 16,
  \dodoi{10.1088/0004-637X/691/1/16}

\bibitem[{{Hoshino}(2015)}]{Hoshino2015}
{Hoshino}, M. 2015, \prl, 114, 061101, \dodoi{10.1103/PhysRevLett.114.061101}

\bibitem[{{Io} \& {Suzuki}(2014)}]{Io2014}
{Io}, Y., \& {Suzuki}, T.~K. 2014, \apj, 780, 46,
  \dodoi{10.1088/0004-637X/780/1/46}

\bibitem[{{Jacquemin-Ide} {et~al.}(2021){Jacquemin-Ide}, {Lesur}, \&
  {Ferreira}}]{Jacquemin-Ide2021}
{Jacquemin-Ide}, J., {Lesur}, G., \& {Ferreira}, J. 2021, \aap, 647, A192,
  \dodoi{10.1051/0004-6361/202039322}

\bibitem[{{Johansen} {et~al.}(2009){Johansen}, {Youdin}, \&
  {Klahr}}]{Johansen2009}
{Johansen}, A., {Youdin}, A., \& {Klahr}, H. 2009, \apj, 697, 1269,
  \dodoi{10.1088/0004-637X/697/2/1269}

\bibitem[{{Kawazura} {et~al.}(2022){Kawazura}, {Schekochihin}, {Barnes},
  {Dorland}, \& {Balbus}}]{Kawazura2022}
{Kawazura}, Y., {Schekochihin}, A.~A., {Barnes}, M., {Dorland}, W., \&
  {Balbus}, S.~A. 2022, Journal of Plasma Physics, 88, 905880311,
  \dodoi{10.1017/S0022377822000460}

\bibitem[{{Kimura} {et~al.}(2016){Kimura}, {Toma}, {Suzuki}, \&
  {Inutsuka}}]{Kimura2016}
{Kimura}, S.~S., {Toma}, K., {Suzuki}, T.~K., \& {Inutsuka}, S.-i. 2016, \apj,
  822, 88, \dodoi{10.3847/0004-637X/822/2/88}

\bibitem[{{Klahr} \& {Bodenheimer}(2003)}]{Klahr03}
{Klahr}, H.~H., \& {Bodenheimer}, P. 2003, \apj, 582, 869,
  \dodoi{10.1086/344743}

\bibitem[{{Kunz} \& {Lesur}(2013)}]{Kunz2013}
{Kunz}, M.~W., \& {Lesur}, G. 2013, \mnras, 434, 2295,
  \dodoi{10.1093/mnras/stt1171}

\bibitem[{{Latter} \& {Papaloizou}(2017)}]{Latter2017}
{Latter}, H.~N., \& {Papaloizou}, J. 2017, \mnras, 472, 1432,
  \dodoi{10.1093/mnras/stx2038}

\bibitem[{{Lazarian} {et~al.}(2020){Lazarian}, {Eyink}, {Jafari}, {Kowal},
  {Li}, {Xu}, \& {Vishniac}}]{Lazarian2020Phpl}
{Lazarian}, A., {Eyink}, G.~L., {Jafari}, A., {et~al.} 2020, Physics of
  Plasmas, 27, 012305, \dodoi{10.1063/1.5110603}

\bibitem[{{Lazarian} \& {Vishniac}(1999)}]{Lazarian1999ApJ}
{Lazarian}, A., \& {Vishniac}, E.~T. 1999, \apj, 517, 700,
  \dodoi{10.1086/307233}

\bibitem[{{Lesur} {et~al.}(2013){Lesur}, {Ferreira}, \& {Ogilvie}}]{Lesur2013}
{Lesur}, G., {Ferreira}, J., \& {Ogilvie}, G.~I. 2013, \aap, 550, A61,
  \dodoi{10.1051/0004-6361/201220395}

\bibitem[{{Lesur} {et~al.}(2022){Lesur}, {Ercolano}, {Flock}, {Lin}, {Yang},
  {Barranco}, {Benitez-Llambay}, {Goodman}, {Johansen}, {Klahr}, {Laibe},
  {Lyra}, {Marcus}, {Nelson}, {Squire}, {Simon}, {Turner}, {Umurhan}, \&
  {Youdin}}]{Lesur2022}
{Lesur}, G., {Ercolano}, B., {Flock}, M., {et~al.} 2022, arXiv e-prints,
  arXiv:2203.09821, \dodoi{10.48550/arXiv.2203.09821}

\bibitem[{{Lightman} \& {Eardley}(1974)}]{Lightman1974}
{Lightman}, A.~P., \& {Eardley}, D.~M. 1974, \apjl, 187, L1,
  \dodoi{10.1086/181377}

\bibitem[{{Machida} {et~al.}(2000){Machida}, {Hayashi}, \&
  {Matsumoto}}]{Machida2000}
{Machida}, M., {Hayashi}, M.~R., \& {Matsumoto}, R. 2000, \apjl, 532, L67,
  \dodoi{10.1086/312553}

\bibitem[{{Masada} {et~al.}(2012){Masada}, {Takiwaki}, {Kotake}, \&
  {Sano}}]{Masada2012ApJ}
{Masada}, Y., {Takiwaki}, T., {Kotake}, K., \& {Sano}, T. 2012, \apj, 759, 110,
  \dodoi{10.1088/0004-637X/759/2/110}

\bibitem[{{Masset}(2000)}]{Masset2000A&AS}
{Masset}, F. 2000, \aaps, 141, 165, \dodoi{10.1051/aas:2000116}

\bibitem[{{Matsumoto} \& {Tajima}(1995)}]{Matsumoto1995}
{Matsumoto}, R., \& {Tajima}, T. 1995, \apj, 445, 767, \dodoi{10.1086/175739}

\bibitem[{{Mishra} {et~al.}(2020){Mishra}, {Begelman}, {Armitage}, \&
  {Simon}}]{Mishra2020}
{Mishra}, B., {Begelman}, M.~C., {Armitage}, P.~J., \& {Simon}, J.~B. 2020,
  \mnras, 492, 1855, \dodoi{10.1093/mnras/stz3572}

\bibitem[{{Mori} {et~al.}(2017){Mori}, {Muranushi}, {Okuzumi}, \&
  {Inutsuka}}]{Mori2017}
{Mori}, S., {Muranushi}, T., {Okuzumi}, S., \& {Inutsuka}, S.-i. 2017, \apj,
  849, 86, \dodoi{10.3847/1538-4357/aa8e42}

\bibitem[{{Narayan} {et~al.}(1987){Narayan}, {Goldreich}, \&
  {Goodman}}]{Narayan1987}
{Narayan}, R., {Goldreich}, P., \& {Goodman}, J. 1987, \mnras, 228, 1,
  \dodoi{10.1093/mnras/228.1.1}

\bibitem[{{Obergaulinger} {et~al.}(2009){Obergaulinger}, {Cerd{\'a}-Dur{\'a}n},
  {M{\"u}ller}, \& {Aloy}}]{Obergaulinger09}
{Obergaulinger}, M., {Cerd{\'a}-Dur{\'a}n}, P., {M{\"u}ller}, E., \& {Aloy},
  M.~A. 2009, \aap, 498, 241, \dodoi{10.1051/0004-6361/200811323}

\bibitem[{{Pucci} {et~al.}(2021){Pucci}, {Tomida}, {Stone}, {Takasao}, {Ji}, \&
  {Okamura}}]{Pucci2021}
{Pucci}, F., {Tomida}, K., {Stone}, J., {et~al.} 2021, \apj, 907, 13,
  \dodoi{10.3847/1538-4357/abc9c0}

\bibitem[{{Riols} \& {Lesur}(2019)}]{Riols2019}
{Riols}, A., \& {Lesur}, G. 2019, \aap, 625, A108,
  \dodoi{10.1051/0004-6361/201834813}

\bibitem[{{Salvesen} {et~al.}(2016){Salvesen}, {Simon}, {Armitage}, \&
  {Begelman}}]{Salvesen2016}
{Salvesen}, G., {Simon}, J.~B., {Armitage}, P.~J., \& {Begelman}, M.~C. 2016,
  \mnras, 457, 857, \dodoi{10.1093/mnras/stw029}

\bibitem[{{Sano} {et~al.}(1999){Sano}, {Inutsuka}, \& {Miyama}}]{Sano99}
{Sano}, T., {Inutsuka}, S., \& {Miyama}, S.~M. 1999, in Astrophysics and Space
  Science Library, Vol. 240, Numerical Astrophysics, ed. S.~M. {Miyama},
  K.~{Tomisaka}, \& T.~{Hanawa} (Boston, MA: Kluwer), 383

\bibitem[{{Sano} \& {Inutsuka}(2001)}]{Sano2001}
{Sano}, T., \& {Inutsuka}, S.-i. 2001, \apjl, 561, L179, \dodoi{10.1086/324763}

\bibitem[{{Scepi} {et~al.}(2018){Scepi}, {Lesur}, {Dubus}, \&
  {Flock}}]{Scepi2018}
{Scepi}, N., {Lesur}, G., {Dubus}, G., \& {Flock}, M. 2018, \aap, 620, A49,
  \dodoi{10.1051/0004-6361/201833921}

\bibitem[{{Shakura} \& {Sunyaev}(1973)}]{Shakura73}
{Shakura}, N.~I., \& {Sunyaev}, R.~A. 1973, \aap, 24, 337

\bibitem[{{Shi} {et~al.}(2016){Shi}, {Stone}, \& {Huang}}]{Shi2016}
{Shi}, J.-M., {Stone}, J.~M., \& {Huang}, C.~X. 2016, \mnras, 456, 2273,
  \dodoi{10.1093/mnras/stv2815}

\bibitem[{{Steinacker} \& {Papaloizou}(2002)}]{Steinacker2002ApJ}
{Steinacker}, A., \& {Papaloizou}, J. C.~B. 2002, \apj, 571, 413,
  \dodoi{10.1086/339892}

\bibitem[{{Stone} {et~al.}(1996){Stone}, {Hawley}, {Gammie}, \&
  {Balbus}}]{Stone1996}
{Stone}, J.~M., {Hawley}, J.~F., {Gammie}, C.~F., \& {Balbus}, S.~A. 1996,
  \apj, 463, 656, \dodoi{10.1086/177280}

\bibitem[{{Suriano} {et~al.}(2019){Suriano}, {Li}, {Krasnopolsky}, {Suzuki}, \&
  {Shang}}]{Suriano2019MNRAS}
{Suriano}, S.~S., {Li}, Z.-Y., {Krasnopolsky}, R., {Suzuki}, T.~K., \& {Shang},
  H. 2019, \mnras, 484, 107, \dodoi{10.1093/mnras/sty3502}

\bibitem[{{Suzuki} \& {Inutsuka}(2009)}]{Suzuki2009ApJ}
{Suzuki}, T.~K., \& {Inutsuka}, S.-i. 2009, \apjl, 691, L49,
  \dodoi{10.1088/0004-637X/691/1/L49}

\bibitem[{{Suzuki} \& {Inutsuka}(2014)}]{Suzuki2014ApJ}
---. 2014, \apj, 784, 121, \dodoi{10.1088/0004-637X/784/2/121}

\bibitem[{{Suzuki} {et~al.}(2010){Suzuki}, {Muto}, \&
  {Inutsuka}}]{Suzuki2010ApJ}
{Suzuki}, T.~K., {Muto}, T., \& {Inutsuka}, S.-i. 2010, \apj, 718, 1289,
  \dodoi{10.1088/0004-637X/718/2/1289}

\bibitem[{{Suzuki} {et~al.}(2019){Suzuki}, {Taki}, \&
  {Suriano}}]{Suzuki2019PASJ}
{Suzuki}, T.~K., {Taki}, T., \& {Suriano}, S.~S. 2019, \pasj, 71, 100,
  \dodoi{10.1093/pasj/psz082}

\bibitem[{{Takasao} {et~al.}(2018){Takasao}, {Tomida}, {Iwasaki}, \&
  {Suzuki}}]{Takasao2018}
{Takasao}, S., {Tomida}, K., {Iwasaki}, K., \& {Suzuki}, T.~K. 2018, \apj, 857,
  4, \dodoi{10.3847/1538-4357/aab5b3}

\bibitem[{{Velikhov}(1959)}]{Velikhov1959}
{Velikhov}, E.~P. 1959, Zh. Eksp. Teor. Fiz., 36, 1398

\bibitem[{{Wissing} {et~al.}(2022){Wissing}, {Shen}, {Wadsley}, \&
  {Quinn}}]{Wissing2022}
{Wissing}, R., {Shen}, S., {Wadsley}, J., \& {Quinn}, T. 2022, \aap, 659, A91,
  \dodoi{10.1051/0004-6361/202141206}

\bibitem[{{Zhu} {et~al.}(2020){Zhu}, {Jiang}, \& {Stone}}]{Zhu2020}
{Zhu}, Z., {Jiang}, Y.-F., \& {Stone}, J.~M. 2020, \mnras, 495, 3494,
  \dodoi{10.1093/mnras/staa952}

\end{thebibliography}
\bibliographystyle{aasjournal}

\end{document}